\documentclass[a4paper]{jpconf}
\usepackage{graphicx,amscd,amsmath,amssymb,verbatim}
\usepackage[pdftex]{hyperref}
\usepackage[TS1,OT1,T1]{fontenc}
\usepackage{mathabx}

\newtheorem{theorem}{Theorem}

\newtheorem{definition}{Definition}

\begin{document}

\title{Relativistic implications of the quantum phase}
\author{Stephen G. Low}
\address{www.Stephen-Low.net}
\ead{Stephen.Low@utexas.edu}
\date{\today}
\begin{abstract}

The quantum phase leads to projective representations of symmetry
groups in quantum mechanics. The projective representations are
equivalent to the unitary representations of the central extension
of the group.\ \ A celebrated example is Wigner's formulation of
special relativistic quantum mechanics as the projective representations
of the inhomogeneous Lorentz group.\ \ However, Wigner's formulation
makes no mention of the Weyl-Heisenberg group and the hermitian
representation of its algebra that are the Heisenberg commutation
relations fundamental to quantum physics. 

We put aside the relativistic symmetry and show that the maximal
quantum symmetry that leaves the Heisenberg commutation relations
invariant is the projective representations of the conformally scaled
inhomogeneous symplectic group.\ \ The Weyl-Heisenberg group and
noncommutative structure arises directly because the quantum phase
requires projective representations.\ \ \ 

We then consider the relativistic implications of the quantum phase
that lead to the Born line element and the projective representations
of an inhomogeneous unitary group that defines a noninertial quantum
theory. (Understanding noninertial quantum mechanics is a prelude
to understanding quantum gravity.) The remarkable properties of
this symmetry and its limits are studied.\ \ 
\end{abstract}
\section{Introduction}

In a well known quote, Dirac states: ``So if one asks what is the
main feature of quantum mechanics, I feel inclined now to say that
it is not noncommutative algebra. It is the existence of probability
amplitudes which underlie all atomic processes. Now a probability
amplitude is related to experiment but only partially. The square
of the modulus is something that we can observe. That is the probability
which the experimental people get. But besides that there is a phase,
a number of modulus unity which we can modify without affecting
the square of the modulus. And this phase is all important because
it is the source of all interference phenomena but its physical
significance is obscure.'' \cite{Dirac 2}

The theme of this paper is to show how this quote from Dirac is
literally the case; the Weyl-Heisenberg group, for which the hermitian
representations of its algebra are the Heisenberg commutation relations,
is a direct consequence of the quantum phase. We will show that
it leads to a symmetry that is of a purely quantum mechanical nature
without any relativistic symmetry being considered. This symmetry
is the automorphisms of the Weyl-Heisenberg group that have the
symplectic group as its most essential subgroup. 

Having obtained this symmetry from purely quantum mechanical considerations,
we then consider its implications on relativity. The immediate observation
is that special relativity, as formulated on spacetime is not the
most general relativistic symmetry that these quantum mechanical
considerations naturally admit.\ \ In fact, as we discuss in this
paper, a more general relativistic symmetry, that includes noninertial
states, naturally presents itself.\ \ This symmetry reduces in the
appropriate inertial limits to include special relativity and, in
the limit of small velocities, nonrelativistic Hamilton's mechanics.
This symmetry, one could say, is the relativistic implications of
the quantum phase.\ \ 
\subsection{Rays and projective representations}

A basic physical property of quantum mechanics is that physical
states are rays.\ \ A ray $\Psi $ is the equivalence class of states
$|\psi \rangle $ in a\ \ Hilbert space $\text{\boldmath $\mathrm{H}$}$
that are defined up to a phase.\ \ 
\begin{equation}
\left. |\psi \right\rangle  \simeq \overset{ }{\left. \left| \widetilde{\psi
}\right. \right\rangle  }\in \Psi \ \ \ \Leftrightarrow \ \ \ \ \overset{
}{\overset{ }{\left. \left| \widetilde{\psi }\right. \right\rangle 
}} =e^{i \omega } \overset{ }{\left. | \psi \right\rangle  }, \left.
\left| \psi \right. \right\rangle  , \overset{ }{\left. \left| \widetilde{\psi
}\right. \right\rangle  }\in \text{\boldmath $\mathrm{H}$}
\end{equation}

\noindent This is a consequence of physical observables in quantum
mechanics being defined by the square of the modulus of the states.
As this square of the modulus has the same value for all state equal
up to a phase, physical observables are defined as the square of
the modulus of the ray to which these states belong.\ \ 
\begin{equation}
{\left| \left( \Phi ,\Psi \right) \right| }^{2}={\left| \left\langle
\phi |\psi \right\rangle   \right| }^{2} ={\left| \left\langle 
\widetilde{\phi }|\widetilde{\psi }\right\rangle   \right| }^{2} .
\end{equation}

 Consider a connected Lie group $\mathcal{G}$with a representation
$\varrho $ acting on the Hilbert space $\text{\boldmath $\mathrm{H}$}$
such that\footnote{We restrict our discussion in this paper to connected\ \ Lie
groups (i.e. all elements are path connected to the identity) as
the central extension is not necessarily unique for a nonconnected
group. Groups that are not connected must be studied on a case by
case basis. }\cite{Azcarraga}
\begin{equation}
\varrho ( g) :\text{\boldmath $\mathrm{H}$}\rightarrow \text{\boldmath
$\mathrm{H}$}:\left. \left. \left| \psi \right. \right\rangle  
\mapsto \left| \psi ^{\prime }\right. \right\rangle  =\varrho (
g) \left. \left| \psi \right. \right\rangle  .
\end{equation}

\noindent If the group and representation is a symmetry of the quantum
state such that 
\begin{equation}
{\left| \left( \varrho ( g) \Phi ,\varrho ( g) \Psi \right) \right|
}^{2}={\left| \left\langle  \varrho ( g) \phi |\varrho ( g) \psi
\right\rangle   \right| }^{2} ={\left| \left\langle  \phi |\psi
\right\rangle   \right| }^{2}\ \ .
\end{equation}

\noindent for all $g\in \mathcal{G}$, then $\mathcal{G}$ is a symmetry
group and $\varrho $ is called a projective representation.\ \ The
cornerstone theorems, (Appendix A, Theorems 1 and 2) state that
a projective representation of a connected Lie group is equivalent
to the linear unitary representation of the group's central extension.
\begin{definition}

The central extension of a connected Lie group $\mathrm{\mathcal{G}}$
is the Lie group $\widecheck{\mathcal{G}}$ that satisfies the following
short exact sequence where $\mathrm{\mathcal{Z}}$ is a maximal abelian
group that is central in $\widecheck{\mathcal{G}}$
\end{definition}
\begin{equation}
\text{\boldmath $e$}\rightarrow \mathcal{Z}\rightarrow \widecheck{\mathcal{G}}\overset{\pi
}{\rightarrow }\mathcal{G}\rightarrow \text{\boldmath $e$} .
\end{equation}

\noindent The abelian group $\mathcal{Z}$ may always be written
as the direct product $\mathcal{Z}\simeq \mathcal{A}( m) \otimes
\mathbb{A}$ of a connected continuous abelian Lie group $\mathcal{A}(
m) \simeq (\mathbb{R}^{m},+)$ and a discrete abelian group $\mathbb{A}$
that may have a finite or countable dimension \cite{bargmann},\cite{mackey2}.

The exact sequence may be decomposed into an exact sequence for
the {\itshape topological} central extension and the {\itshape algebraic}
central extension,
\begin{equation}
\text{\boldmath $e$}\rightarrow \mathbb{A}\rightarrow \overline{\mathcal{G}}\overset{\pi
\mbox{}^{\circ}}{\rightarrow }\mathcal{G}\rightarrow \text{\boldmath
$e$} \text{\boldmath $,$}\text{\boldmath $\ \ $}\text{\boldmath
$e$}\rightarrow \mathcal{A}( m) \rightarrow \widecheck{\mathcal{G}}\overset{\widetilde{\pi
}}{\rightarrow }\overline{\mathcal{G}}\rightarrow \text{\boldmath
$e$}.
\end{equation}

\noindent where $\pi =\pi \mbox{}^{\circ}\circ \widetilde{\pi }$. The
first exact sequence defines the universal cover where $\mathbb{A}\simeq
\ker  \pi \mbox{}^{\circ}$ is the finite abelian fundamental\ \ homotopy
group. 

All of the groups is in the second sequence are simply connected
and therefore may be defined by the exponential map of the central
extension of its Lie algebra. Therefore, it is necessary to determine
the central extension of the Lie algebra of the group.

An algebraic central extension of a Lie algebra $g$ is the Lie algebra
$\widecheck{g}$ that satisfies the following short exact sequence where
$z$ is the maximal abelian algebra that is central in $\widecheck{g}$,
\begin{equation}
\text{\boldmath $0$}\rightarrow z\rightarrow \widecheck{g}\rightarrow
g\rightarrow \text{\boldmath $0$} .
\end{equation}

\noindent where $\text{\boldmath $0$}$ is the trivial algebra. Suppose
$\{X_{a}\}$ is a basis of the Lie algebra $g$ with commutation relations
$[X_{a},X_{b}]=c_{a,b}^{c}X_{c}$, $a,b=1,...r$.\ \ Then an algebraic
central extension is a maximal set of central abelian generators
$\{A_{\alpha }\}$, where $\alpha ,\beta ,... =1,..m$,\ \ such that
\begin{equation}
\left[ A_{\alpha },A_{\beta }\right] =0,\ \ \ \ \left[ X_{a},A_{\alpha
}\right] =0,\ \ \ \ \left[ X_{a},X_{b}\right] =c_{a,b}^{c}X_{c}+c_{a,b}^{\alpha
}A_{\alpha }.
\end{equation}

\noindent The basis $\{X_{a},A_{\alpha }\}$ of the centrally extended
Lie algebra must also satisfy the Jacobi identities. The Jacobi
identities constrain the admissible central extensions of the algebra.
The choice\ \ $X_{a}\mapsto X_{a}+A_{a}$ will always satisfy these
relations and this trivial case is excluded.\ \ The algebra $\widecheck{g}$
constructed in this manner is equivalent to the central extension
of $g$ given in Definition 1.

 The full central extension may therefore be computed by determining
the universal covering group of the algebraic central extension.\ \ 

Levi's theorem (Appendix A, Theorem 4)\ \ states that any simply
connected Lie group may be written as the semidirect product of
a simply connected semi-simple Lie group and a simply connected
solvable Lie group\footnote{Of course, either the semi-simple or
solvable subgroup may be trivial.}. As a central extension of a
group is always simply connected, it may always be written in this
semidirect product form. Furthermore, Mackey's theorems \cite{mackey},\cite{Low12}
gives the prescription with which to compute the unitary representations
of a general class of semi-direct product groups. 

The application of these theorems given in Appendix A and the above
comments on the computation of the central extension enables us
to compute the projective representations of a general class of
connected Lie groups.\footnote{There are some technical considerations
for the Mackey theorem's to apply. All of the Lie groups studied
in this paper satisfy the conditions.}\ \ 
\subsection{Projective representations of the inhomogeneous Lorentz
group: Special relativistic quantum mechanics.}

The connected inhomogeneous Lorentz group is 
\begin{equation}
\mathcal{I}\mathcal{L}( 1,n) \simeq \mathcal{L}( 1,n) \otimes _{s}\mathcal{A}(
n+1) .
\end{equation}

\noindent The physical case is $n=3$. $\mathcal{L}( 1,n) $ is the
proper Lorentz group that is the connected component of $\mathcal{O}(
1,n) $.\footnote{We are considering only the connected component
to avoid technical issues with the central extension mentioned previously
that are not essential to this paper. The discrete\ \ parity-time
group associated with the four components of $\mathcal{O}( 1,n)
$ are key to the full theory.} The Lorentz group leaves invariant
the Minkowski line element
\begin{equation}
d \tau ^{2}=\eta _{a,b}d x^{a}d x^{b}=d t^{2}-\frac{1}{c^{2}}d q{}^{2}.%
\label{rrr: Minkowski line element}
\end{equation}

\noindent $\mathcal{A}( n) $ is the abelian translation group that
is isomorphic to $\mathbb{R}^{n}$ regarded as a Lie group under
addition, $\mathcal{A}( n) \simeq (\mathbb{R}^{n},+)$. The inhomogeneous
Lorentz group is a matrix group and it elements $\Gamma \in \mathcal{I}\mathcal{L}(
1, n) $ may be realized by $n+2$ dimensional matrices.\ \ 
\begin{equation}
\Gamma ( \Lambda ,a) =\left( \begin{array}{ll}
 \Lambda  & a \\
 0 & 1
\end{array}\right) ,\ \ \ \Lambda \in \mathcal{L}( 1,n) , a\in \mathbb{R}^{n+1}.
\end{equation}

\noindent where $\Lambda ^{\mathrm{t}}\eta  \Lambda =\eta $, det
$\Lambda $=1 and $\Lambda _{0,0}>1$ and $\eta $ is the $n+1$ dimensional
matrix for the Minkowski metric,\ \ $\eta =\operatorname{diag}(
1,-1,...,-1) $.

For the quantum theory, projective representations of the inhomogeneous
Lorentz group are required because physical states are rays \cite{wigner},\cite{Weinberg1}.
The projective representations are calculated as the unitary representations
of its central extension using the cornerstone theorem (Appendix
A, Theorem 2). 

It turns out that for the inhomogeneous Lorentz group that we cannot
add any algebraic central generators without violating the Jacobi
identities and so the central extension, the Poincar\'e group, is
just the cover, $\widecheck{\mathcal{I}\mathcal{L}}( 1,n) \simeq \overline{\mathcal{I}\mathcal{L}}(
1,n) $\ \ [Weinberg1],
\begin{equation}
\mathcal{P}( 1,n) \simeq \overline{\mathcal{I}\mathcal{L}}( 1,n)
\simeq \overline{\mathcal{L}}( 1,n) \otimes _{s}\mathcal{A}( n+1)
.%
\label{rrr: Poincare group}
\end{equation}

\noindent For $n=3$,\ \ $\overline{\mathcal{L}}( 1,3) \simeq \mathcal{S}\mathcal{L}(
2,\mathbb{C}) $.\ \ The Mackey theorems may then be used to compute
the unitary irreducible representations \cite{Weinberg1}.\ \ For
our purposes here, we note only that for the faithful representations
for the timelike case, the little group is $\mathcal{S}\mathcal{U}(
2) $ and that the Hilbert space for the timelike representation
is $\text{\boldmath $\mathrm{H}$}={\text{\boldmath $L$}}^{2}( \mathbb{H}^{+},\mathbb{V}^{2s+1})
$ where\ \ $\mathbb{H}^{+}\simeq \mathcal{S}\mathcal{L}( 2,\mathbb{C})
/\mathcal{S}\mathcal{U}( 2) $. $\mathbb{V}^{2s+1}$ is a $2s+1$ dimensional
complex vector space for the ordinary unitary representations of
the little group, $\mathcal{S}\mathcal{U}( n) $ for which $s$ is
the spin eigenvalue, $s=0,\frac{1}{2},1,...$. 

The Poincar\'e group does not contain the Weyl-Heisenberg group
as a subgroup. The representations of the algebra of the Weyl-Heisenberg
group are the Heisenberg commutation relations.\ \ These relations,
that are fundamental to quantum mechanics do not appear in the unitary
representations of the Poincar\'e group as prescribed by the Mackey
theorems. They must be added on in an essentially {\itshape ad hoc
}manner after the fact. 
\subsection{Projective representations of the inhomogeneous Euclidean
group: Galilean relativistic quantum mechanics.}

The nonrelativistic (or more correctly, the Galilean relativity)
limit refers to the limit of special relativity for small velocities
relative to the speed of light, $\frac{v}{c}\rightarrow 0$. This
is mathematically modeled by the limit of $c\rightarrow \infty $.\ \ Under
this limit, the Special Relativistic (SR) Minkowski line element
(10) reduces to the Galilean Relativistic (GaR) line element $d
{\tau \mbox{}^{\circ}}^{2}=d t^{2}$\ \ that is Newtonian {\itshape
absolute} time.\ \ 

The connected invariance group for the degenerate line element $d
t^{2}$ is the affine group $\mathcal{I}\mathcal{G}\mathcal{L}( n,\mathbb{R})
$\cite{Glimore2},\cite{Low7}
\begin{equation}
\begin{array}{lll}
 \mathrm{SR} &   & \mathrm{GaR} \\
 d \tau ^{2}  & \operatorname*{\rightarrow }\limits_{c \rightarrow
\infty } & d t^{2} \\
 \mathcal{L}( 1,n)  &   & \mathcal{I}\mathcal{G}\mathcal{L}( n,\mathbb{R})
\end{array}%
\label{rrr. SR GaR groups}
\end{equation}

\noindent The abelian subgroup of the affine symmetry is parameterized
by velocity 
\begin{equation}
\mathcal{I}\mathcal{G}\mathcal{L}( n,\mathbb{R}) \simeq \mathcal{G}\mathcal{L}(
n,\mathbb{R}) \otimes _{s}\mathcal{A}( n) 
\end{equation}

It reduces to the Euclidean group if we also require invariance
of length in the rest frame.\ \ This affine symmetry requires only
the definition Newtonian invariant time $d t^{2}$ and does not require
any knowledge of the special relativistic line element for which
it is a limit\footnote{It is easy to show that for the connected
symmetry group, this is equivalent to the usual affine definition
that requires invariance of $d t$}.\ \ 

On the other hand, the Lorentz group $\mathcal{L}( 1,n) $ contracts
to the Euclidean group $\mathcal{E}( n) \simeq \mathcal{S}\mathcal{O}(
n) \otimes _{s}\mathcal{A}( n) $ that is parameterized by rotations
on nonrelativistic velocity.\ \ The inhomogeneous Lorentz group
likewise contracts to the inhomogeneous Euclidean group $\mathcal{I}\mathcal{E}(
n) \simeq \mathcal{E}( n) \otimes _{s}\mathcal{A}( n+1) $ \cite{inonu}.
\begin{equation}
\begin{array}{lll}
 \mathrm{SR} &   & \mathrm{GaR} \\
 \mathcal{L}( 1, n)   & \operatorname*{\rightarrow }\limits_{c \rightarrow
\infty } & \mathcal{E}( n)  \\
 \mathcal{I}\mathcal{L}( 1,n)  & \operatorname*{\rightarrow }\limits_{c
\rightarrow \infty } & \mathcal{I}\mathcal{E}( n) 
\end{array}%
\label{rrr: Lorentz contraction}
\end{equation}

The Galilean relativistic quantum theory requires the projective
representations of the inhomogeneous Euclidean group $\mathcal{I}\mathcal{E}(
n) $.\ \ The nonzero commutators of the Lie algebra of the inhomogeneous
Euclidean group are 
\begin{equation}
\begin{array}{l}
 \left[ J_{i,j},J_{k,l}\right] =J_{j,k} \delta _{i,l}+J_{i,l} \delta
_{j,k}-J_{i,k} \delta _{j,l}-J_{j,l} \delta _{i,k}, \\
 \left[ J_{i,j},G_{k}\right] =G_{j} \delta _{i,k}-{\widehat{G}}_{i}\delta
_{j,k},\ \ \left[ J_{i,j},P_{k}\right] =P_{i} \delta _{j,k}-P_{j}\delta
_{i,k}, \\
 \left[ G_{i},E\right] =P_{i}.
\end{array}%
\label{rrr: Ga algebra unitary rep}
\end{equation}

\noindent This algebra admits a one dimensional algebraic extension
with a generator $M$ that is physically interpreted as mass
\begin{equation}
 \left[ G_{i},P_{k}\right] =M \delta _{i,k}.%
\label{rrr: Ga M}
\end{equation}

\noindent The group with this centrally extended algebra is the
Galilei group $\mathcal{G}a( n) $,
\begin{equation}
\mathcal{G}a( n) \simeq \mathcal{E}( n) \otimes _{s}\left( \mathcal{A}(
n+1) \otimes \mathcal{A}( 1) \right) .%
\label{rrr: Poincare group}
\end{equation}

\noindent The cover of the Galilei group is the central extension
of the inhomogeneous Euclidean group,
\begin{equation}
\widecheck{\mathcal{I}\mathcal{E}}( n) \simeq \overline{\mathcal{G}a}(
n) .
\end{equation}

\noindent For $n=3$,\ \ $\overline{\mathcal{G}a}( 3) \simeq \mathcal{S}\mathcal{U}(
2) \otimes _{s}\mathcal{A}( 3) \otimes _{s}\mathcal{A}( 5) $.\ \ 

The Mackey theorems may then be used to compute the unitary irreducible
representations \cite{Low12}\cite{Voisin}\cite{inonu2,wikramasekara}.\ \ For
our purposes here, we note only that, for $n=3$, that the faithful
representations of the timelike case, the little group is $\mathcal{S}\mathcal{U}(
2) $ and that the Hilbert space for the timelike representation
is $\text{\boldmath $\mathrm{H}$}={\text{\boldmath $L$}}^{2}( \mathbb{H}^{+},\mathbb{V}^{2s+1})
$ where\ \ $\mathbb{H}^{+}\simeq \overline{\mathcal{E}}( 3) /\mathcal{S}\mathcal{U}(
2) \simeq \mathbb{R}^{3}$.\ \ $\mathbb{V}^{2s+1}$ is a $2s+1$ dimensional
complex vector space for the ordinary unitary representations of
the little group, $\mathcal{S}\mathcal{U}( n) $ for which $s$ is
the spin eigenvalue, $ s=0,\frac{1}{2},1,...$. 

The Galilei group does not contain the Weyl-Heisenberg group as
a subgroup. Again, these relations, that are fundamental to quantum
mechanics, do not appear in the unitary representations of the Galilei
group as prescribed by the Mackey theorems. They must again be added
on in an essentially {\itshape ad hoc }manner after the fact. 
\section{Quantum symmetry of Heisenberg commutation relations }

In this section, we put aside the Minkowski metric and relativistic
considerations and consider only the inhomogeneous symplectic symmetry
on a phase space $\mathbb{P}\simeq \mathbb{R}^{2n}$ that is well
known from the Hamilton formulation of classical mechanics,\ \ 
\begin{equation}
\mathcal{I}\mathcal{S}p( 2n) \simeq \mathcal{S}p( 2n) \otimes _{s}\mathcal{A}(
2n) %
\label{rrr: ISp definition}
\end{equation}

\noindent $\mathcal{I}\mathcal{S}p( 2n) $ is a matrix group realized
with $2n+1$ dimensional matrices\ \ 
\begin{equation}
\Gamma ( \mathrm{A},z) =\left( \begin{array}{ll}
 {\mathrm A} & z \\
 0 & 1
\end{array}\right) , \mathrm{A}\in \mathcal{S}p( 2n) , z\in \mathbb{R}^{2n}\text{}
\end{equation}

\noindent The abelian group $\mathcal{A}( 2n) $ are translations
of points in the phase space,\ \ $z\in \mathbb{P}\mbox{}^{\circ}\simeq
\mathbb{R}^{2n}$.\ \ These points may be written as $z=(p,q)$ with
$p,q\in \mathbb{R}^{n}$ that are momentum and position degrees of
freedom. The symplectic group acts on the tangent space $T_{z}\mathbb{P}\mbox{}^{\circ}$
or cotangent space $T_{z}^{*}\mathbb{P}\mbox{}^{\circ}$.\ \ The
corresponding quantum symmetry is the projective representation
of the inhomogeneous symplectic group. Again, by the fundamental
theorem (Appendix A, Theorem 2), these are equivalent to the unitary
representations of its central extension. This central extension
may be computed by determining the central extension of the Lie
algebra of the group and constructing the simply connected group
with this algebra.\ \ This calculation results in 
\begin{equation}
\mathcal{I}\widecheck{\mathcal{S}p}( 2n) \simeq \overline{\mathcal{S}p}(
2n) \otimes _{s}\mathcal{H}( n) .%
\label{rrr: ISP CE definition}
\end{equation}

This satisfies Levi's theorem (Appendix A, Theorem 4) with a simply
connected, semi-simple homogeneous group $\overline{\mathcal{S}p}(
2n) $ and the simply connected solvable normal subgroup $\mathcal{H}(
n) $ that is the Weyl-Heisenberg group. The Weyl-Heisenberg group
is a one parameter central extension of the abelian group $\mathcal{A}(
2n) $.\ \ 

This is not the most general central extension of the abelian group.
The Lie algebra of the abelian group is $[A_{\alpha },A_{\beta }]=0$,
$\alpha ,\beta ,...=1,...,2n$.\ \ The central extension of the algebra
is 
\begin{equation}
\left[ A_{\alpha },A_{\beta }\right] =M_{\alpha ,\beta }, \left[
M_{\alpha ,\beta },M_{\gamma ,\kappa }\right] =0,\ \ \ \left[ A_{\alpha
},M_{\gamma ,\kappa }\right] =0,\ \ 
\end{equation}

\noindent with $M_{\alpha ,\beta }=-M_{\beta .\alpha }$ and therefore
have $n( 2n-1) $ generators.\ \ It follows that the central extension
of the abelian group, $\widecheck{\mathcal{A}}( 2n) $ is $n( 2n-1) $
dimensional.\ \ However, if the abelian group is a subgroup of the
inhomogeneous symplectic group, the Jacobi identities that result
from the algebra of the homogeneous symplectic group constrains
its central extension to be precisely the Weyl-Heisenberg group.
\subsection{The Weyl-Heisenberg group}

The Weyl-Heisenberg group is a semi-direct product group,
\begin{equation}
\mathcal{H}( n) \simeq \mathcal{A}( n) \otimes _{s}\mathcal{A}(
n+1) .
\end{equation}

\noindent It is a real matrix group,\ \ $p,q\in \mathbb{R}^{n},\ \ \text{$\iota$$
\in $$ \mathbb{R}$}$ where the $(2n+2)\times (2n+2)$ matrix realization
is [Major],[Low4]
\begin{equation}
\Upsilon ( p,q,\iota ) =\left( \begin{array}{llll}
 1_{n} & 0 & 0 & q \\
 0 & 1_{n} & 0 & p \\
 p^{\mathrm{t}} & -q^{\mathrm{t}} & 1 & 2 \iota  \\
 0 & 0 & 0 & 1
\end{array}\right) .%
\label{rrr:Heisenberg matrix group}
\end{equation}

The group product and inverse are given by matrix multiplication
\begin{gather}
\Upsilon ( p^{\prime },q^{\prime },\iota ^{\prime }) \Upsilon (
p,q,\iota ) =\Upsilon ( p^{\prime }+p,q^{\prime }+q,\iota +\iota
^{\prime }+\frac{1}{2}\left( p^{\prime }\cdot q-q^{\prime }\cdot
p\right) ) ,
\\\Upsilon ^{-1}( p,q,\iota ) =\Upsilon ^{-1}( -p,-q,-\iota ) .
\end{gather}

\noindent It immediately follows that $\Upsilon ( 0,q,\iota ) \in
\mathcal{A}( n+1) $ is a normal abelian subgroup and $\Upsilon (
p,0,0) \in \mathcal{A}( n) $ is an abelian homogeneous subgroup
leading to the semidirect product structure. This is not the only
semidirect product structure. One can show equally that $\Upsilon
( p,0,\iota ) \in \mathcal{A}( n+1) $ is a normal abelian subgroup
and $\Upsilon ( 0,q,0) \in \mathcal{A}( n) $ is an abelian homogenous
subgroup also leading to a semidirect product structure.

The Lie algebra is easily determined by differentiating the matrix
representatives and evaluating at the identity. A general element
of the matrix algebra is [Low5]
\begin{equation}
Z=\frac{1}{\hbar }\left( q^{i}P_{i}+p^{i}Q_{i}\right) +i I=\left(
\begin{array}{llll}
 0 & 0 & 0 & q \\
 0 & 0 & 0 & p \\
 p^{\mathrm{t}} & -q^{\mathrm{t}} & 0 & 2 \iota  \\
 0 & 0 & 0 & 0
\end{array}\right) .%
\label{rrr:Heisenberg matrix algebra}
\end{equation}

\noindent This satisfies the Heisenberg algebra
\begin{equation}
\left[ P_{i},Q_{j}\right] =\hbar  \delta _{i,j}I
\end{equation}

\noindent with $[A,B]=A B-B A$ and\ \ $i,j,..=1,...,n$.
\subsection{Unitary representations of the Weyl-Heisenberg group}

The unitary representations result can be calculated from the Mackey
theorems as the Weyl-Heisenberg group is a semidirect product group.\ \ Using
the semidirect product form with the normal subgroup parameterized
by position results in the faithful unitary representation $\varrho
$ with position diagonal results in
\begin{equation}
\left( \varrho ( \Upsilon ( p,q,\iota ) ) \psi \right) \left( \widetilde{q}\right)
=e^{i \left( \nu  \left(  \iota  +\frac{p\cdot q}{2\hbar }\right)
+\frac{p\cdot \widetilde{q}}{\hbar }\right) }\psi ( \widetilde{q}+q) ,
\end{equation}

\noindent where $\psi \in {\text{\boldmath $L$}}^{2}( \mathbb{R}^{n},\mathbb{C})
$ and $\Upsilon ( p, q, \iota ) \in \mathcal{H}( n) $\footnote{The
Mackey theorem constructs the unitary irreducible representations
where-as the Stone-von Neumann theorem proves that, once obtained,
they are indeed the unitary irreducible representations} \cite{Major},\cite{Low4}.
The hermitian representation ${\widehat{P}}_{i}=\varrho ^{\prime }(
P_{i}) $, ${\widehat{Q}}_{i}=\varrho ^{\prime }( Q_{i}) $, $\widehat{I}=\varrho
^{\prime }( I) $ of the algebra is 
\begin{equation}
\left\langle  q\right| {\widehat{P}}_{i}\left| \psi \right\rangle  =i
\nu  \hbar  \frac{\partial }{\partial  q^{i}}\psi ( q) , \left\langle
q\right| {\widehat{Q}}_{i}\left| \psi \right\rangle  =q_{i}\psi ( q)
, \left\langle  q\right| \widehat{I}\left| \psi \right\rangle  =\nu
\psi ( q) ,%
\label{rrr: q Heisenberg generator rep}
\end{equation}

\noindent where $\nu \in \mathbb{R}\backslash \{0\}.$ If instead
we use the semidirect product form with the normal subgroup parameterized
by momentum, then this results in faithful representations with
momentum diagonal, 
\begin{equation}
\left( \varrho ( \Upsilon ( p,q,\iota ) ) \psi \right) \left( \widetilde{p}\right)
=e^{i \left( \nu  \left(  \iota  -\frac{p\cdot q}{2\hbar }\right)
+\frac{q\cdot \widetilde{p}}{\hbar }\right) }\psi ( \widetilde{p}+p) 
\end{equation}

\noindent and in this case the hermitian representation of the algebra
takes the form 
\begin{equation}
\left\langle  p\right| {\widehat{P}}_{i}\left| \psi \right\rangle  =p_{i},
\left\langle  p\right| {\widehat{Q}}_{i}\left| \psi \right\rangle  =-i
\hbar  \frac{\partial }{\partial  q^{i}}\psi ( p) , \left\langle
p\right| \widehat{I}\left| \psi \right\rangle  =\nu  \psi ( p) .%
\label{rrr: p Heisenberg generator rep}
\end{equation}

\noindent The generators in either (31) or (33) satisfy the Heisenberg
Lie algebra relations $[{\widehat{P}}_{i},{\widehat{Q}}_{j}]=i \nu  \hbar
\delta _{i,j}\widehat{I}$. By Schur's lemma, $\widehat{I}$ is the identity
and, choosing $\nu =1$, these are the Heisenberg commutation relations
\begin{equation}
\left[ {\widehat{P}}_{i},{\widehat{Q}}_{j}\right] =i \hbar  \delta _{i,j}.
\end{equation}

We pause here to recap.\ \ We started with the very basic fact that,
in quantum mechanics, physical states are rays with measurable quantities
given by the square of their modulus.\ \ Quantum symmetries are
the projective representation of the group.\ \ With a homogenous
symplectic symmetry, this results in the abelian group becoming
the nonabelian Weyl-Heisenberg group; the `noncommutative structure'
in Dirac's quote at the beginning of this paper.\ \ 
\subsection{The Weyl-Heisenberg automorphism group}

We turn now to the question of determining the maximal symmetry
with the Weyl-Heisenberg group as a normal subgroup. This enables
us to determine the maximal quantum symmetry that admits Heisenberg
commutation relations. Again, at this point there is no line element
defining relativistic concepts such as how time transforms between
the physical states. 

Theorem 5 (Appendix A) states that the maximal group that contains
a group $\mathcal{G}$ is its automorphism group ${\mathcal{A}ut}_{\mathcal{G}}$.\ \ The
Weyl-Heisenberg group is a matrix group and we first determine the
maximal group of automorphisms that are also a matrix group.\ \ These
must preserve the form of the matrix group elements, 
\begin{equation}
\Upsilon ( z^{\prime },\iota ^{\prime }) =\varsigma _{\Omega }\Upsilon
( Z,\iota )  =\Omega  \Upsilon ( Z,\iota )  \Omega ^{-1},
\end{equation}

\noindent where $z=(p,q)\in \mathbb{R}^{2n}$ and\ \ $\Upsilon (
z,\iota ) ,\Upsilon ( z^{\prime },\iota ^{\prime }) \in \mathcal{H}(
n) $.\ \ \ Expanding this out using (25) with $z=(q,p)$\ \ this
is
\begin{equation}
\left( \begin{array}{lll}
 1_{2n} & 0 & z^{\prime } \\
 {z^{\prime }}^{\mathrm{t}}\zeta \mbox{}^{\circ} & 1 & 2 \iota ^{\prime
} \\
 0 & 0 & 1
\end{array}\right) =\Omega \left( \begin{array}{lll}
 1_{2n} & 0 & z \\
 z^{\mathrm{t}}\zeta \mbox{}^{\circ} & 1 & 2 \iota  \\
 0 & 0 & 1
\end{array}\right) \Omega ^{-1},\ \ \zeta \mbox{}^{\circ}=\left(
\begin{array}{ll}
 0 & 1_{n} \\
 -1_{n} & 0
\end{array}\right) .
\end{equation}

\noindent Then a simple matrix calculation shows that the general
form of $\Omega $ is \cite{Low9}, \cite{Low8}, \cite{folland}
\begin{equation}
\Omega  =\left( \begin{array}{lll}
 \delta  A & 0 & z \\
 z^{\mathrm{t}}\zeta \mbox{}^{\circ} A & \delta ^{2}\epsilon  &
\iota  \\
 0 & 0 & \epsilon 
\end{array}\right) ,
\end{equation}

\noindent where $A\in \mathcal{S}p( 2n) $, $z=\{q,p\}\in \mathbb{R}^{2n}$,
$\iota \in \mathbb{R}$, $\delta \in \mathbb{R}\backslash \{0\}$
and $\epsilon =\pm 1$.

An analysis of the matrix realization for $\Omega $ shows that these
matrices are elements of a matrix group that is given by
\begin{equation}
{\mathrm{aut}}_{\mathcal{H}( n) }\simeq \left( \mathbb{Z}_{2,2}\otimes
_{s}\mathcal{D}\right) \otimes _{s}\mathcal{H}\mathcal{S}p( 2n)
,\ \ \ \mathcal{H}\mathcal{S}p( 2n) =\mathcal{S}p( 2n) \otimes _{s}\mathcal{H}(
n) .
\end{equation}

\noindent where $\mathcal{D}\simeq (\mathbb{R}^{+},\times )$ and
$\mathbb{Z}_{2,2}\simeq \mathbb{Z}_{2}\otimes \mathbb{Z}_{2}$. We
confine ourselves from this point on to consider the connected component\ \ $\mathcal{D}\otimes
_{s}\mathcal{H}\mathcal{S}p( 2n) $ of the group ${\mathrm{aut}}_{\mathcal{H}(
n) }$.\ \ As an automorphism group is maximally centrally extended
(Appendix A, Theorem 4), the connected component of the automorphism
group is the central extension of this group is
\begin{equation}
{\mathcal{A}ut}_{\mathcal{H}( n) }\simeq {\widecheck{\mathrm{aut}}}_{\mathcal{H}(
n) }\simeq \mathcal{D}\otimes _{s}\overline{\mathcal{S}p}( 2n) \otimes
_{s}\mathcal{H}( n) .
\end{equation}

\noindent A calculation shows that this may be written in terms
as the central extension of a group $\mathcal{D}\mathcal{S}p( 2n)
$,\ \ 
\begin{equation}
{\mathcal{A}ut}_{\mathcal{H}( n) }\simeq \mathcal{D}\widecheck{\mathcal{S}p}(
2n) ,%
\label{rrr: DSp automorphism defn}
\end{equation}

\noindent where 
\begin{equation}
\mathcal{D}\mathcal{S}p( 2n) \simeq \text{}\mathcal{D}\otimes _{s}\mathcal{I}\mathcal{S}p(
2n) , 
\end{equation}

\noindent and $\mathcal{I}\mathcal{S}p( 2n) $ is given by (20).
A computation also shows that
\begin{equation}
\mathcal{D}\widecheck{\mathcal{S}p}( 2n) \simeq \mathcal{D}\otimes _{s}\mathcal{I}\widecheck{\mathcal{S}p}(
2n) .%
\label{rrr: DSP CE factors}
\end{equation}

\noindent where $\widecheck{\mathcal{I}\mathcal{S}p}( 2n) $ is given
in (22).

This gives us the key result that the maximal connected group that
has a Weyl-Heisenberg normal subgroup is the central extension of
$\mathcal{D}\mathcal{S}p( 2n) $.\ \ The one dimensional conformal
scale subgroup $\mathcal{D}$ appears in a semidirect product with
$\mathcal{I}\mathcal{S}p( 2n) $ that is the basic classical mechanics
symmetry. The central extension of the conformally scaled inhomogeneous
symplectic group $\mathcal{D}\mathcal{S}p( 2n) $ has a Weyl-Heisenberg
subgroup that is the central extension of its abelian subgroup constrained
by the symplectic homogenous group. The connected automorphism group
is defined in terms of the symplectic group as given in (42). This
shows the fundamental relationship between the symplectic group
and the Weyl-Heisenberg group. It is only possible to construct
a connected semidirect product with a Weyl-Heisenberg subgroup $\mathcal{K}\otimes
_{s}\mathcal{H}( n) $ if $\mathcal{K}\subset \mathcal{D}\otimes
_{s}\mathcal{S}p( 2n) $.\ \ For example, semidirect products with
$\mathcal{K}\simeq \mathcal{G}\mathcal{L}( 2n,\mathbb{R}) $ or $\mathcal{S}\mathcal{O}(
2n) $ do not exists as these are not subgroups as required. 
\subsection{Maximal quantum symmetry group with Heisenberg commutation
relations}

Quantum symmetries require us to consider projective representations
of the connected automorphism group.\ \ By the fundamental theorem,
these representations are equivalent to the unitary representations
of its central extension.\ \ An automorphism group is its own central
extension. The automorphism group of the Weyl-Heisenberg group is
equivalent to the central extension of $\mathcal{D}\mathcal{S}p(
2n) $. The projective representations of $\mathcal{D}\mathcal{S}p(
2n) $ are therefore equivalent to the unitary representations of
the connected component of the automorphism group ${\mathcal{A}ut}_{\mathcal{H}(
n) }$ of the Weyl-Heisenberg group $\mathcal{H}( n) $.\ \ \ The
automorphism group ${\mathcal{A}ut}_{\mathcal{H}( n) }$ is the largest
group that has a Weyl-Heisenberg normal subgroup.\ \ The hermitian
representation of its algebra, corresponding to the unitary representations
of the\ \ group,\ \ are the Heisenberg commutation relations.\ \ Furthermore,
this group is the largest connected symmetry group that leaves invariant
the Weyl-Heisenberg normal subgroup and its algebra invariant.\ \ Finally,
the unitary representations defines the Hilbert space on which the
group and algebra act, and in particular the Weyl-Heisenberg subgroup
and subalgebra act. 

In short, one can say that the group $\mathcal{D}\mathcal{S}p( 2n)
\simeq \mathcal{D}\otimes _{s}\mathcal{I}\mathcal{S}p( 2n) $ defines
the maximally connected symmetry group whose projective representations,
required by quantum mechanics, leave the Heisenberg commutation
relations invariant.\ \ 

This is of fundamental physical importance.\ \ Consider a unitary
operator $\widehat{U}=\varrho ( g) $ that is a unitary representation
$\varrho $ of an element $g\in \mathcal{G}$ where $\mathcal{G}$
is a symmetry or relativity group. The\ \ unitary operator transforms
states of the Hilbert space of the representation,\ \ $|\psi \rangle
\in \text{\boldmath $\mathrm{H}$}$
\begin{equation}
\widetilde{\left. \left| \psi \right. \right\rangle  }=\widehat{U}\left|
\psi \right\rangle  
\end{equation}

\noindent and the Hermitian operators that realize the algebra transform
as 
\begin{equation}
{\widehat{X}}^{\prime }=\widehat{U}\widehat{X}{\widehat{U}}^{-1}.
\end{equation}

\noindent In particular, for the Weyl-Heisenberg group, 
\begin{equation}
{{\widehat{P}}^{\prime }}_{i}=\widehat{U}{\widehat{P}}_{i}{\widehat{U}}^{-1},\ \ {{\widehat{Q}}^{\prime
}}_{i}=\widehat{U}{\widehat{Q}}_{i}{\widehat{U}}^{-1},\ \ {\widehat{I}}^{\prime
}=\widehat{U}\widehat{I}{\widehat{U}}^{-1}=\widehat{I}.
\end{equation}

Now, we want the Heisenberg commutation relations to be preserved
under the transformation so that the uncertainty principle is valid
in all states in the Hilbert related by the unitary transformation
$\widehat{U}$.\ \ This requires that\ \ 
\begin{equation}
i \hbar  \delta _{i,j}{\widehat{I}}^{\prime } =\left[  {{\widehat{P}}^{\prime
}}_{i},{{\widehat{Q}}^{\prime }}_{j}\right] =\widehat{U}[  {\widehat{P}}_{i},{\widehat{Q}}_{j}]
{\widehat{U}}^{-1}= i \hbar  \delta _{i,j}\widehat{U}\widehat{I}{\widehat{U}}^{-1}=i
\hbar  \delta _{i,j}\widehat{I}.
\end{equation}

\noindent Otherwise, we would be able to transform to points in
the Hilbert space where the Heisenberg commutation relations, and
hence at these points in the Hilbert space,\ \ the quantum uncertainty
principle does not hold!

As the representation is faithful, this is true if and only if\ \ the
group and algebra elements satisfies
\begin{equation}
\left[  {P^{\prime }}_{i},{Q^{\prime }}_{j}\right] =g[  P_{i},Q_{j}]
g^{-1}.%
\label{rrr: Heisenberg pq commuation relations}
\end{equation}

\noindent This means that $g$ is an element of the automorphism
group of the Weyl-Heisenberg group and algebra\footnote{The automorphism
group for a group and its algebra in general may be different. For
the Weyl-Heisenberg group they are the same. }. 

The above arguments may also be extended to extended phases space
$\mathbb{P}\simeq \mathbb{R}^{2n+2}$ that includes time and energy
degrees of freedom. At this point we have not introduced any relativistic
structure to distinguish time and energy from momentum and position.
There is no line element defining invariant time, null cones or
any concepts of past and future. Note also that neither the symplectic
group nor the Weyl-Heisenberg group have any concept of invariant
signature\ \ that is a property of the Lorentz group (and more generally,
that any real orthogonal group $\mathcal{O}( m,n) $).\ \ At this
point, these two degrees of freedom are just two additional degrees
of freedom in the symplectic manifold with a symmetry that is now
$\mathcal{D}\mathcal{S}p( 2n+2) \simeq \mathcal{D}\otimes _{s}\mathcal{I}\mathcal{S}p(
2n+2) $ where
\begin{equation}
\mathcal{I}\mathcal{S}p( 2n+2) \simeq \mathcal{S}p( 2n+2) \otimes
_{s}\mathcal{A}( 2n+2) .
\end{equation}

The time energy Heisenberg commutation relations that result from
the projective representations are, at least formally,
\begin{equation}
\left[ \widehat{T},\widehat{E}\right] =i \widehat{I}.
\end{equation}

These also, together with the momentum-position commutation relations
(47),\ \ are invariant under the projective representations of $\mathcal{D}\mathcal{S}p(
2n+2) $ in exactly the same manner as described above. 

The projective representations of $\mathcal{D}\mathcal{S}p( 2n+2)
$ define the representations and Hilbert space for the most general
quantum mechanics in which the Heisenberg commutation relations
hold\footnote{We are confining ourselves to connected groups. The
most general group turns out to be $\mathbb{Z}_{2,2}\otimes _{s}\mathcal{D}\mathcal{S}p(
2n+2) $}.\ \ Its central extension is the connected component of
the automorphism group\footnote{Both the Poincar\'e and Galilei
groups are subgroups of\ \ ${\mathcal{A}ut}_{\mathcal{H}( n+1) }$.}
\begin{equation}
{\mathcal{A}ut}_{\mathcal{H}( n+1) }\simeq \mathcal{D}\otimes _{s}\widecheck{\mathcal{I}\mathcal{S}p}(
2n+2) \simeq \mathcal{D}\otimes _{s}\overline{\mathcal{S}p}( 2n+2)
\otimes _{s}\mathcal{H}( n+1) .
\end{equation}

\noindent The Mackey theorems show that these representations result
in a Hilbert space ${\text{\boldmath $L$}}^{2}( \mathbb{R}^{n+1},\text{\boldmath
$\mathrm{H}$}) $ so that the wave functions are functions of position
and time $\psi ( t,q) $ or any of the other commuting subsets of
the Weyl-Heisenberg group such as\ \ $\psi ( e,p) ,\psi ( t,p) $
$\mathrm{or} $$\psi ( e,q) $ and not all the degrees of phase space
together. 
\section{Reciprocal relativity of noninertial states}

In the introduction, we briefly reviewed the special relativistic
inhomogeneous Lorentz group, $\mathcal{I}\mathcal{L}( 1,n) $, symmetry
of classical (i.e. non quantum mechanical) special relativity.\ \ We
then reviewed that the basis of special relativistic quantum mechanics
follows simply by considering the projective representations of
the inhomogeneous Lorentz group that are required by the quantum
phase \cite{wigner},\cite{Weinberg1}.\ \ By the fundament theorem
(Appendix A, Theorem 2), these are equivalent to the unitary representations
of its central extension, the Poincar\'e group.\ \ \ These same
considerations apply in the non-relativistic limit that mathematically
is $c\rightarrow \infty $.\ \ The inhomogeneous group contracts
to the classical inhomogeneous Euclidean group $\mathcal{I}\mathcal{E}(
n) $ that describes the relativistic symmetry of classical nonrelativistic
Newtonian mechanics.\ \ Again, a quantum formulation follows by
simply considering the projective representations of $\mathcal{I}\mathcal{E}(
n) $ that are equivalent to the unitary representations of its central
extension, the Galilei group. This central extension admits a new
algebraic generator that is nonrelativistic mass. 

However, neither of these formulations make any reference the Weyl-Heisenberg
group and the Heisenberg commutation relations\ \ that are foundational
to quantum mechanics. 

In the second chapter, we put aside the relativistic metric and
considered only the Weyl-Heisenberg group.\ \ It is a particular
central extension of the abelian group $\mathcal{A}( 2n) $ of translations
on phase space. Therefore, the unitary representations are a particular
projective representation of $\mathcal{A}( 2n ) $ corresponding
to a one dimensional central extension.\ \ The Hilbert space for
these unitary representations is the Hilbert space that we are expecting
for elementary non-relativistic quantum mechanics.\ \ Furthermore,
we show that the maximal connected quantum symmetry that preserves
the Heisenberg commutation relations are the projective representations
of $\mathcal{D}\mathcal{S}p( 2n) $. $\mathcal{D}\mathcal{S}p( 2n)
$ is a classical (i.e. non-quantum) symmetry on phase space that
is just $\mathcal{I}\mathcal{S}p( 2n) $ that is very familiar from
elementary classical Hamiltonian mechanics with an additional conformal
scaling group $\mathcal{D}$. The addition of two more degrees of
freedom to the phase space and symmetry, $\mathcal{D}\mathcal{S}p(
2n+2) $ and again, this is a classical symmetry on extended phase
space.\ \ 

The symmetry $\mathcal{D}\mathcal{S}p( 2n+2) $ on extended phases
space $\mathbb{P}$ resulted without any relativistic consideration.\ \ There
is no invariant time line element. In the remainder of the paper,
we combine these two ideas by adding back the relativistic orthogonal
line element to this symmetry $\mathcal{D}\mathcal{S}p( 2n+2) $.\ \ In
a sense we are `{\itshape relativizing} the quantum symmetry' rather
than the usual `quantizing of a relativistic symmetry'. We require
that the symmetry be the subgroup of the maximal quantum phase $\mathcal{D}\mathcal{S}p(
2n+2) $ on extended phase space $\mathbb{P}\simeq \mathbb{R}^{2n+2}$
that leaves invariant the relativistic line element defining proper
time. 

The $\mathcal{D}\mathcal{S}p( 2n+2) $ symmetry is a classical symmetry
on extended phase space. In particular, its homogeneous group is
$\mathcal{D}\otimes \mathcal{S}p( 2n) $.\ \ In this section we study
the non-quantum classical consequences of the symmetry that, for
reasons that we will shortly discuss, refer to as reciprocal relativity.
Then, in Section 4 we will study the projective representations
of the resulting groups to study the corresponding quantum theories.
\subsection{The relativistic line elements }

The natural choice for proper time to consider first is the Minkowski
line element\ \ \ 
\begin{equation}
d \tau ^{2}=\eta _{a,b}d x^{a}d x^{b}=d t^{2}-\frac{1}{c^{2}}d q{}^{2}
\end{equation}

\noindent The Minkowski line elements is a degenerate line element
on extended phase space. Using the co-ordinates\ \ $\{z^{\alpha
}\}\simeq \{x^{a},y^{a}\}$ $\alpha ,\beta =1,...2n+2$, $a,b=0,..n$,
the line element may be written in terms of the degenerate metric
on extended phase space, 
\begin{equation}
d \tau ^{2}={\widetilde{\eta }}_{\alpha ,\beta }d z^{\alpha }d z^{\beta
},\ \ \ \ \widetilde{\eta }= \left( \begin{array}{ll}
 \eta  & 0 \\
 0 & 0
\end{array}\right) .%
\label{rrr: Minkowski line element}
\end{equation}

We have encountered a degenerate line element previously.\ \ \ Recall
that the degenerate line element $d t^{2}$ on Galilean space time
defines invariant Newtonian time (13).\ \ It is a result of taking
the $c\rightarrow \infty $ limit of the special relativistic Minkowski
line element.\ \ 

This suggests that the degenerate Minkowski line element on phase
space is itself the limit of a nondegenerate line element on extended
phase space. This leads us to consider a nondegenerate orthogonal
metric on the extended phase space for which the Minkowski and Newtonian
line elements are limiting forms, just as the Newtonian element
is a limit of the Minkowski metric on spacetime.\ \ 

We therefore postulate the nondegenerate Born line element on extended
phase space \cite{born1},\cite{born2},\cite{low11},\cite{Low3},\cite{Low5},\cite{jarvis-1,morgan}
\begin{equation}
d s^{2}= d t^{2}-\frac{1}{c^{2}}d q{}^{2}+\frac{1}{b^{2}}\left(
\frac{1}{c^{2}}d \varepsilon {}^{2} -d p^{2}\right) .%
\label{rrr: Born line element}
\end{equation}

\noindent From dimensional analysis, the constant $b$ must have
dimensions of force,(i.e. in MKS units, Newtons). We will show shortly
that it can be taken to be one of the three fundamental dimensional
scales along with $c$ and $\hbar $. In the limit $b\rightarrow \infty
$, the Born line element contracts to the Minkowski line element
(10).\ \ \ 

This line element was suggested by Born as a consequence of his
reciprocity principle \cite{born1},\cite{born2}. This reciprocity
principle is a discrete symmetry that stated that physics should
be invariant under the discrete transform $q\rightarrow p, p\rightarrow
-q$. Born had no concept of a continuous group of symmetries. However,
in recognition of Born's initial insight, we refer to this theory
that has the Born line element as reciprocal relativity. 
\subsection{Reciprocal relativity: Time dilation}

The Born metric\ \ (6)\ \ may be written as
\begin{equation}
\begin{array}{ll}
 d s^{2} & =d t^{2}-\frac{1}{c^{2}}d q{}^{2}-\frac{1}{b^{2}}d p^{2}+\frac{1}{c^{2}b^{2}}d
\varepsilon {}^{2} \\
  & =d t^{2}( 1-\frac{1}{c^{2}}v{}^{2}-\frac{1}{b^{2}}f^{2}+\frac{1}{c^{2}b^{2}}r{}^{2})
\end{array}%
\label{rrr. Born metric}
\end{equation}

\noindent where 
\begin{equation}
\mathrm{velocity}: v^{i}=\frac{d q^{i}}{d t},\ \ \mathrm{force}:
f^{i}=\frac{d p^{i}}{d t},\ \ \mathrm{power}: r=\frac{d \varepsilon
}{d t}.%
\label{rrr: force power velocity}
\end{equation}

\noindent The velocity, force and power are not constants but are
functions of position, momentum and time, $v^{i}( p,q,t) $, $f^{i}(
p,q,t) $ and $r( p,q,t) \text{}$. 

The time dilation formula that defines relative rate that the clocks
tick depends on the relative noninertial motion of the state
\begin{equation}
\begin{array}{ll}
 d t & =\frac{1}{\sqrt{1-\frac{v{}^{2}}{c^{2}}-\frac{f^{2}}{b^{2}}+\frac{r{}^{2}}{c^{2}b^{2}}}}
d s.
\end{array}%
\label{rrrr: time dilation}
\end{equation}

\noindent For the inertial states where $f=0$, $r=0$. this clearly
reduces to the usual expression from special relativity
\begin{equation}
\begin{array}{ll}
 d t & =\frac{1}{\sqrt{1-\frac{v{}^{2}}{c^{2}}}} d s
\end{array}.%
\label{rrr: sr time dilation}
\end{equation}

The special relativistic time dilation depends only on the relative
inertial state of the particles; the reciprocal relativistic expression
(56) depends on the relative noninertial states.\ \ In special relativity,
there is a global inertial state that all observers (that must be
in a physical state) agree on where-as the rest state is relative
to each physical state.\ \ In the reciprocal relativity, both the
rest state and the inertial state are relative to the physical states.\ \ There
is no absolute inertial frame as there is in special relativity.\ \ Just
as special relativity abolishes the notion of the absolute rest
frame, reciprocal relativity abolishes also the notion of an absolute
inertial frame.\ \ In special relativity, the effects that the rest
frame is relative and not absolute are only apparent for velocities
such that $v/c$ is significant. As $v/c\rightarrow 0$, we have the
illusion that the rest frame is absolute in Galilean relativity.\ \ This
is the case also with reciprocal relativity; as $f/b\rightarrow
0$ and $r/b c\rightarrow 0$, we have the illusion that the inertial
frame is absolute.\ \ That the inertial frame is not absolute is
only observable in regimes that $f/b$ and/or $r/b c$ are significant.\ \ If
$b$ is large, as discussed in the following section, this may be
in a regime that has either not yet or is just now being probed
experimentally. 

The Born metric may be written in {\itshape four} notation as the
sum of Minkowski proper time (3) and the differential of the mass
line element on energy-momentum space\ \ 
\begin{equation}
d s^{2}=d \tau ^{2}+\frac{c^{2}}{b^{2}} d \mu ^{2}%
\label{rrr: born metric proper time and mass}
\end{equation}

\noindent where $d \mu ^{2}$ is the differential of the mass that
is the energy-momentum line element 
\begin{equation}
{c }^{2}d \mu ^{2}=\eta _{a,b} d p^{a}d p^{b}=\frac{1}{c^{2}} d
e^{2}-d p^{2}.%
\label{rrr: mass line element}
\end{equation}

\noindent Using the definitions of force and power in (9) and the
definition of the mass line element $ d \mu ^{2} $ in (12) it follows
that 
\begin{equation}
\begin{array}{ll}
 {c^{2}( \frac{d \mu }{d t}) }^{2} & =
\end{array}\frac{1}{c^{2}}r{}^{2}-f^{2}
\end{equation}

\noindent and therefore the time dilation may be written
\begin{equation}
\begin{array}{ll}
 d t & =\frac{1}{\sqrt{1-\frac{v{}^{2}}{c^{2}}-\frac{c^{2}}{b^{2}}{\left(
\frac{d \mu }{d t}\right) }^{2}}} d s.
\end{array}
\end{equation}

This means that the time dilation can be understood in terms of
the rate at which the state is moving off the mass shell.\ \ It
is difficult to find phenomena in the classical regime where this
occurs. However, this clearly does occur in the quantum regime.
The quantum theory that we outline in the next chapter, is again
just the projective representations of the classical symmetry. 

In special relativity, the concepts of proper time $d \tau ^{2}$\ \ and
the differential of mas $d \mu ^{2}$ are independent invariants.\ \ In
fact, if we require both of these to be independently invariant,\ \ then
the subgroup of $\mathcal{D}\otimes \mathcal{S}p( 2n) $ that leaves
them invariant is $\mathcal{L}( 1,n) $.\ \ This is the inertial
subgroup that constrains the state to the special relativistic mass
shell.\ \ In the noninertial theory, these are combined to define
the Born metric (12)\ \ and now special relativistic proper time
$d \tau $ is no longer an invariant of the theory. The special relativity
proper time $d \tau $ is related to the reciprocal relativity invariant
time $d s$ as the state moves off the mass shell 
\begin{equation}
\begin{array}{ll}
 d \tau  & =\frac{1}{\sqrt{1+\frac{c^{2}}{b^{2}}{\left( \frac{d
\mu {}}{d \tau }\right) }^{2}}}d s.
\end{array}
\end{equation}
\subsection{Doesn't General Relativity address noninertial states?}

A question that is almost always asked when stating that reciprocal
relativity addresses noninertial states is that ``Doesn't General
Relativity address noninertial states and therefore hasn't this
question already been answered''.\ \ Let us review carefully what
special and general relativity do address.\ \ 

Special relativity addresses the transformation between two inertial
physical states using the inhomogeneous Lorentz group. This leads
to the well known time dilation of special relativity (57) and the
other effects such as length contraction, null hypersurfaces, the
causal null cones and so forth. 

 The equivalence principle of general relativity states that a freely
falling state under the influence of gravity is locally equivalent
to an inertial state. This leads to a theory in which an `apparent'
noninertial states due to the `force' of gravity is actually a locally
inertial state on a curved spacetime. In a purely gravitating system,
all particles follow geodesics that are locally inertial trajectories
in this curved space-time. Neighboring clocks are related by the
line element ${\mathrm{d\tau }}^{2}={g( x) }_{\mu ,\nu }d x^{\mu
}d x^{\nu }$ where the metric ${g( x) }_{\mu ,\nu }$ now depends
on the location $x$ in space-time.\ \ The connection translates
vectors between neighboring locally inertial frames and the covariant
derivative is relative to the locally inertial frame. The dynamics
are now in the\ \ field equations for curvature that then determine
the metric and connection. 

However,\ \ neither of these principles directly address the noninertial
state of a particle that is due to one of the other forces that
is not gravity, say an electron in a electromagnetic field. One
cannot directly encompass this in the general relativistic gravitational
theory such that the electrodynamic forces also become part of the
geometry. The noninertial state of the electron cannot be geometrized
into a locally inertial state on a curved manifold as general relativity
description of gravity does. We have tried for almost a 100 years,
starting with Kaluza,\ \ but to date this has not been successful.\ \ 

Furthermore, the principle of special relativity does not directly
address either the transformations defining time dilation between
an inertial state and this noninertial state of the electron in
the magnetic field.\ \ In strict special relativity, the Lorentz
transformations hold between inertial states. It is a separate assumption
that the Lorentz transformation, and the corresponding time dilation
(57) continues to hold in this case where one of the physical states
is following a noninertial trajectory such as the electron on the
electromagnetic field.\ \ It is an assumption that the usual special
relativity time dilation continues to hold. Clearly if the noninertial
state is `nearly inertial' with respect to some force scale, then
the special relativistic equations hold. This is clearly stated
in Einstein's original paper on special relativity \cite{einstein}.
However, whether or not it is the case for physical states that
are strongly noninertial with respect to some scale is an empirical
question that is not determined a {\itshape priori}.\ \ 

The noninertial quantum symmetry that is given by the projective
representations of $\mathcal{D}\mathcal{S}p( 2n+2) $ is a symmetry
on extended phase space. (We assert that understanding noninertial
symmetry in quantum mechanics is a prelude to quantum gravity.)
The Minkowski line element is degenerate on extended phase space.
The Born metric is the natural nondegenerate line element on extended
phase space that generalizes the Minkowski line element. It introduces
the force scale $b$ and as $f/b\rightarrow 0$ and $r/b c \rightarrow
0$, it reduces to special relativity\ \ as an approximation. 
\subsection{Planck scales}

We have noted that the constant $b$ in the Born line element is
a universal constant with the dimensions of force.\ \ Usually, we
take $\{c,G,\hbar \}$ as the three dimensionally independent scales
and define the Planck scales of time, length, momentum and energy
$\{\lambda _{t},\lambda _{q}, \lambda _{p},\lambda _{\varepsilon
}\}$ as
\begin{equation}
\lambda _{t}=\sqrt{\frac{G \hbar }{ c^{5}}},\ \ \lambda _{q}=\sqrt{\frac{\hbar
G}{c^{3} }},\ \ \lambda _{p}=\sqrt{\frac{\hbar  c^{3}}{G }},\ \ \lambda
_{\varepsilon }=\sqrt{\frac{\hbar \ \ c^{5}}{G} }.%
\label{rrr: Planck scales G}
\end{equation}

\noindent More fundamental are the set of three identities that
relate the four fundamental scales $\{\lambda _{t},\lambda _{q},\lambda
_{p},\lambda _{\varepsilon }\}$ to three fundamental scales $\{c,b,\hbar
\}$:
\begin{equation}
\frac{\lambda _{q}}{\lambda _{t}}=c=\frac{\lambda _{\varepsilon
}}{\lambda _{p}},\ \ \ \ \ \lambda _{q} \lambda _{p}=\hbar =\lambda
_{t}\lambda _{\varepsilon },\ \ \ \frac{\lambda _{p}}{\lambda _{t}}=b=\frac{\lambda
_{\varepsilon }}{\lambda _{q}}%
\label{rrr: fundamental scale relations}
\end{equation}

\noindent The constant $b$ has the dimensions of force (in MKS units,
Newtons). A very straightforward calculations shows that these relations
result in the usual scales given in (63) if and only if 
\begin{equation}
b=\frac{c^{4}}{G}%
\label{rrr: b equation}
\end{equation}

\noindent In this expression $G$ is taken to be the basic dimensional
constant and $b$ a convenient derived constant.\ \ 

The constant $b$ that has the dimensions of force that is precisely
the constant that appears in the Born line element.\ \ This, and
the simple canonical form of (64) suggests that we turn the equation
(65) around and instead write\ \ \ \ 
\begin{equation}
G=\alpha _{G}\frac{c^{4}}{b}.
\end{equation}

\noindent Now $G$ is a derived constant and its content has been
absorbed into the dimensionless coupling constant $\alpha _{G}$.\ \ $b$
is now the third fundamental constant in terms of which the Planck
dimension scales become
\begin{equation}
\lambda _{t}=\sqrt{\frac{ \hbar }{b c}},\ \ \lambda _{q}=\sqrt{\frac{\hbar
c}{b }},\ \ \lambda _{p}=\sqrt{\frac{\hbar  b}{c }},\ \ \lambda
_{\varepsilon }=\sqrt{\hbar  b c }.%
\label{rrr: Planck scales b}
\end{equation}

 If $\alpha _{G}=1$, these are numerically identical to (63) and
this is just notation. However, we make the hypothesis that is not
the case and $\alpha _{G}$ may have some other value that must be
determined experimentally.\ \ \ 

It is also useful to define the scales with $c=1$,\ \ for use with
{\itshape four} notation, 
\begin{equation}
\lambda _{x}=\sqrt{\frac{ \hbar }{b }},\ \ \ \ \lambda _{p}=\sqrt{\hbar
b }.%
\label{rrr: Planck scales b}
\end{equation}

The Born metric and the constant $b$ are the only new postulates
in this theory and $\alpha _{G}$ is the only free parameter. Whether
$b$ is actually a fundamental dimensional scale and whether the
Born line element is realized by nature is a matter of empirical
verification.\ \ $\alpha _{G}$ sets the scales at which the reciprocally
relativistic effects manifest and it is probably somewhere between
the weak interaction energy scale and the usual Planck energy given
in (20).\ \ This gives $\alpha _{G}$ a range of ${10}^{-17}$ to
1.\ \ 

The mathematical formulation is always in terms of dimensionless
entities.\ \ General dimensionless co-ordinates $z^{\alpha }$ of
extended phase space may be written in terms of dimensioned time,
position, momentum and energy coordinates as $\{z^{\alpha }\}=\{\frac{1}{\lambda
_{t}}t,\frac{1}{\lambda _{q}}q^{i},\frac{1}{\lambda _{p}}p^{i},\frac{1}{\lambda
_{\varepsilon }}\varepsilon \}$, $i=1,...,n$.\ \ This completely
sets the scaling with respect to the three basic dimensional scale
constants $\{c,b,\hbar \}$.\ \ For example, the line element is
\begin{equation}
\begin{array}{rl}
 d s^{2} & =\eta _{\alpha ,\beta }d z^{\alpha }d z^{\beta }=\frac{-1}{\lambda
_{t}^{2}}d t^{2}+\frac{1}{\lambda _{q}^{2}}d q^{2}+\frac{1}{\lambda
_{p}^{2}}d p^{2}-\frac{1}{\lambda _{\varepsilon }^{2}}d \varepsilon
^{2} \\
  & =\frac{-1}{\lambda _{t}^{2}}\left(  d t^{2}-\frac{1}{c^{2}}d
q{}^{2}-\frac{1}{b^{2}}d p^{2}+\frac{1}{b^{2}c^{2}}d \varepsilon
{}^{2} \right) .
\end{array}
\end{equation}

\noindent This is the scaling given in (54).\ \ Another example
is the symplectic metric, 
\begin{equation}
\begin{array}{rl}
 d s^{2} & =\zeta _{\alpha ,\beta }d z^{\alpha }d z^{\beta }=-\frac{1}{\lambda
_{\varepsilon }}d \varepsilon \wedge \frac{1}{\lambda _{t}}d t +
\frac{1}{\lambda _{p}}d p_{i} \wedge \frac{1}{\lambda _{q}}d q^{i}
\\
  & =\frac{1}{\hbar }\left(  -d \varepsilon \wedge d t + d p_{i}
\wedge d q^{i}\right) .
\end{array}
\end{equation}

\noindent Again, this is the expected scaling, and one finds that
this basic scaling completely determines the scaling of the theory
with respect to the three basic dimensional scale constants.\ \ 
\subsection{The null hypersurface}

The null hypersurface defines essential properties of the relativistic
theory.\ \ It is a fixed point surface so that if a particle state
is on the null surface, classically it is not possible for it to
leave the surface.\ \ \ For the usual special relativistic theory
on space time, the null hypersurface is defined by the line element
(10) having a value of zero 
\begin{equation}
0=1-\frac{v{}^{2}}{c^{2}},%
\label{rrr: inertial light cone}
\end{equation}

\noindent and so $v=\pm c$.\ \ This defines the null cones $\frac{d
q{}^{2}}{c^{2}} = d t$ that, at a point in the manifold $\mathbb{M}=\mathbb{R}^{n+1}$,
define the differential light cones on $T_{(t,q)}^{*}\mathbb{M}$
that defines the past and future regions $d \tau ^{2}>0$ as well
as the spacelike regimes $d \tau ^{2}<0$\ \ and the null hypersurface
$d \tau ^{2}=0$.\ \ \ \ 

In reciprocal relativity, the states are in general noninertial
with the above inertial case as a special case. The null hypersurface
depends not only on the rate of change of position, (velocity $v=\frac{d
q{}}{d t}$) but also depends the rate of change of momentum, (force
$f=\frac{d p{}}{d t}$), and the rate of change of energy, (power
$r=\frac{d \varepsilon {}}{d t}$).\ \ 

Setting $d s^{2}=0$ in the\ \ Born line element (54) gives the null
hypersurface condition\ \ 
\begin{equation}
0=1-\frac{v{}^{2}}{c^{2}}-\frac{f^{2}}{b^{2}}+\frac{r{}^{2}}{c^{2}b^{2}}%
\label{rrr: power force velocity}
\end{equation}

\noindent or equivalently
\begin{equation}
\frac{v{}^{2}}{c^{2}}+\frac{f^{2}}{b^{2}}=1+\frac{r{}^{2}}{c^{2}b^{2}}.%
\label{rrr: noninertial light cone}
\end{equation}

\noindent This reduces to the inertial case (71) when $f=r=0$ and
therefore the inertial null surface is the $d q, d t$ plane slice\ \ of
the null surface with velocity constant, {\itshape v=$ \pm $c}.
As states become noninertial and move off the inertial plane, velocity
is no longer constant. 

\noindent The differential null cones are 
\begin{equation}
\frac{d q{}^{2}}{c^{2}}+\frac{d p^{2}}{b^{2}}=d t^{2}\left( 1+\frac{r{}^{2}}{c^{2}b^{2}}\right)
.%
\label{rrr: null cones}
\end{equation}

\noindent In the case $n=1$, the extended phase space is 4 dimensional
and the null hypersurface is 3 dimensional.\ \ One can choose to
plot the $\{d t,d q, d p\}$ dimensions and to parameterize the $\{d
\varepsilon \}$ dimension. This results in elliptical null cones
that\ \ flatten with increased $r$ as the term on the right hand
side of (24) becomes larger.\ \ 

As in the case of special relativity, these cones define timelike
future and past with $d s^{2}>0$,\ \ spacelike regions with $d s^{2}<0$
and the null surface itself with $d s^{2}=0$.\ \ The usual notions
of causality are extended to this space with the null hypersurfaces
constraining the velocity, force and power for each of the regions.\ \ Again,
the null hypersurface is a fixed point surface and therefore there
is no way in the classical theory for states to move between the
spacelike and timelike regions. Furthermore, states on the null
hypersurface must remain on the hypersurface in the non-quantum
approximation.\ \ 
\begin{equation}
v{}=\pm c\sqrt{1-\frac{f^{2}}{b^{2}}+\frac{r{}^{2}}{c^{2}b^{2}}}=\pm
c\sqrt{1+\frac{c^{2}}{b^{2}}{\left( \frac{d \mu {}}{d t}\right)
}^{2}}.
\end{equation}

Consider some example points on the null surface. For $f=0$, $r=0$,
$v=\pm c$\ \ but at another point on the hypersurface, $f=\pm b$,
$r=0$ and therefore $v=0$.\ \ The null surface also has values for
which velocity is greater than $c$.\ \ For example,\ \ $f=0$, $r=2
b c$ and so $v=\pm 2 c$.\ \ In fact, there are points on the null
hypersurface for which $c$ takes values from $0$ to $\pm \infty
$. These points have a value greater than $c$ provided that the
mass changes in flight,\ \ $\frac{d \mu {}}{d t}\neq 0$.\ \ 

The case $v>c$\ \ for light has not yet been amenable for laboratory
experiment\footnote{The OPERA experiment has recently given preliminary
results showing neutrinos with a velocity greater than $c$. These
neutrinos have three flavors with slightly different masses. The
flavors are known to change in flight leading to a $\frac{d \mu
}{d t}>0$. Whether is a possible test will be the topic of a future
paper.} \cite{OPERA}. Yet there is strong evidence that in the very
early universe that this was the case during the inflationary epoch.
During this time, light was probing extreme noninertial regimes
of the null hypersurface for which $r \gg b c$ .
\subsection{Reciprocal relativity symmetry group}

One condition for the homogeneous classical symmetry for reciprocal
relativity is that it leaves the Born line element invariant. The
symmetry group of this line element that acts on the tangent and
cotangent spaces of extended phase space is, of course, $\mathcal{O}(
2,2n) \subset \mathcal{G}\mathcal{L}( 2n+2,\mathbb{R}) $.\ \ In
addition it must be a subgroup of the $\mathcal{D}\mathcal{S}p(
2n+2) $ symmetry in order to leave the Heisenberg commutation relations
invariant when we consider the projective representations required
by quantum mechanics. Its homogeneous symmetry is $\mathcal{D}\otimes
\mathcal{S}p( 2n+2) $ that is also a subgroup of $\mathcal{G}\mathcal{L}(
2n+2,\mathbb{R}) $ and also acts on the tangent and cotangent space
of extended phase space. The group satisfying both of these conditions
is the intersection of these two groups \cite{Low3}
\begin{equation}
\mathcal{O}( 2,2n) \cap \mathcal{D}\otimes \mathcal{S}p( 2n+2) \simeq
\mathcal{U}( 1,n)  \subset \mathcal{G}\mathcal{L}( 2n+2,\mathbb{R})
.
\end{equation}

The group $\mathcal{U}( 1,n) \simeq \mathcal{U}( 1) \otimes \mathcal{S}\mathcal{U}(
1,n) $ is a matrix group for which elements can be realized by $2n+2$
dimensional matrices\ \ $\Gamma ( \theta ,\Lambda ,\mathrm{M}) =e^{i
\theta }\Gamma ( \Lambda ,\mathrm{M}) $\ \ where $e^{\mathrm{i\theta
}}\in \mathcal{U}( 1) $ and $\Gamma ( \Lambda ,\mathrm{M}) \in \mathcal{S}\mathcal{U}(
1,n) $,
\begin{equation}
\Gamma ( \Lambda ,\mathrm{M}) =\Delta \left( \begin{array}{ll}
 \Lambda  & {\mathrm{M}}^{\mathrm{t}} \\
 {\mathrm M} & \Lambda 
\end{array}\right) ,\ \ \Delta ={\left( \mathrm{Det}\left( \begin{array}{ll}
 \Lambda  & {\mathrm{M}}^{\mathrm{t}} \\
 {\mathrm M} & \Lambda 
\end{array}\right) \right) }^{-1},%
\label{rrr: U Lambda M}
\end{equation}

\noindent where $\Lambda $ are $2n+1$ dimensional matrices that
have the property $\Lambda  \eta  \Lambda ^{\mathrm{t}}=\eta $,
$\mathrm{Det} \Lambda =1$ such that $\Lambda \in \mathcal{L}( 1,n)
$.\ \ The $\mathrm{M}$ are $2n+1$ dimensional matrices satisfying
the symmetric property that ${\mathrm{M}}^{\mathrm{t}}=\eta  \mathrm{M}
\eta $. Therefore, $\mathrm{Det} \Gamma ( \Lambda ,\mathrm{M}) =1$
as required for it to be an element of $\mathcal{S}\mathcal{U}(
1,n) $.

For an extended phase space $\mathbb{P}\simeq \mathbb{R}^{2n+2}$,
we also have translation invariance, $\mathcal{A}( 2n+2) $ so that
the full inhomogeneous symmetry is 
\begin{equation}
\mathcal{I}\mathcal{U}( 1,n)  \simeq \mathcal{U}( 1, n) \otimes
_{s}\mathcal{A}( 2n+2) .
\end{equation}
\subsection{Invariance groups for the contracted line elements}

The Minkowski line element $d \tau ^{2}$ on spacetime of standard
special relativity contracts to invariant Newtonian time line element
$d t^{2}$ in the limit of $c\rightarrow \infty $,\ \ 
\begin{equation}
d \tau ^{2}= d t^{2}-\frac{1}{c^{2}} d q^{2}\operatorname*{\rightarrow
}\limits_{c\rightarrow \infty } d t^{2}.
\end{equation}

This limit determines the effective theory in the limit of small\ \ rates
of change of position, velocity, $v/c\rightarrow 0 $. The connected
invariance group of the Minkowski line element of Special Relativity
(SR) is the Lorentz group $\mathcal{L}( 1,n) $ and the connected
invariance group of the Newtonian time line element of Galilean
Relativity (GaR) is the affine group $\mathcal{I}\mathcal{G}\mathcal{L}(
n,\mathbb{R}) $ as given in (13).

For the line element for Reciprocal Relativity (RR) on extended
phase space, we have corresponding contractions $c\rightarrow \infty
$ corresponding to small velocities and also $b\rightarrow \infty
$ corresponding to small rates of change of momentum and energy;
that is, small force $f/\mathrm{cb}\rightarrow 0$ and power $r/b
c\rightarrow 0$.\ \ The Born line element $d s^{2}$ (54) on extended
phase space contracts to the Minkowski line element $d \tau ^{2}$
in the limit of $b\rightarrow \infty $ that in turn contracts to
the Newtonian line element. The corresponding connected homogenous
groups that leave these line elements invariant are 
\begin{equation}
\begin{array}{lllll}
 \mathrm{RR} &   & \mathrm{SR} &   & \mathrm{GaR} \\
 d s^{2} & \operatorname*{\rightarrow }\limits_{b \rightarrow \infty
} & d \tau ^{2}  & \operatorname*{\rightarrow }\limits_{c \rightarrow
\infty } & d t^{2} \\
 \mathcal{U}( 1,n)  &   & \mathcal{L}a( 1,n)   &   & \mathcal{H}\mathcal{S}p(
2n) 
\end{array}
\end{equation}

\noindent The $\mathcal{L}a( 1,n) $ group is defined by \cite{Low9}
\begin{equation}
\mathcal{L}a( 1,n) =\mathcal{L}( 1,n) \otimes _{s}\mathcal{A}( m)
,\ \ m=\frac{\left( n+1\right) \left( n+2\right) }{2}\text{},%
\label{rrr: La definition}
\end{equation}

\noindent and the group $\mathcal{H}\mathcal{S}p( 2n) $ is the semidirect
product of the symplectic group with the Weyl-Heisenberg group,\ \ \ 
\begin{equation}
\mathcal{H}\mathcal{S}p( 2n) =\mathcal{S}p( 2n) \otimes _{s}\mathcal{H}(
n) .
\end{equation}

The group $\mathcal{L}a( 1,n) $ and its physical consequences is
studied in \cite{Low9}.\ \ We note here only that\ \ is a matrix
group with elements that may be realized by $2n+2$ dimensional matrices
that are, with the basis ordering $\{d t, d q^{i}, d \varepsilon
,d p^{i}\}$,\ \ \ \ \ 
\begin{equation}
\Gamma ( \Lambda ,\mathrm{M\mbox{}^{\circ}}) =\left( \begin{array}{ll}
 \Lambda  & 0 \\
 \mathrm{M\mbox{}^{\circ}} & \Lambda 
\end{array}\right) \in \mathcal{L}a( n) .\ \ %
\label{rrr: la matrix}
\end{equation}

\noindent The $\Lambda \in \mathcal{L}( 1,n) $ are $n+1$ dimensional
submatrices that may be parameterized as usual by rotations and
velocity $v^{i}/c$.\ \ The abelian normal subgroup is given in terms
of the $2n+1$ submatrices $\mathrm{M\mbox{}^{\circ}}$ that are traceless
and satisfy the symmetry condition,\ \ ${\mathrm{M\mbox{}^{\circ}}}^{\mathrm{t}}=\eta
\mathrm{M\mbox{}^{\circ}} \eta $.\ \ The abelian subgroup is parameterized
by force $f^{i}/b$, power $r/b c$ and stress $m^{i,j}/b c$ [25].
The algebra of the $\mathcal{L}a( 1,n) $ group is given in (126).

The group $\mathcal{H}\mathcal{S}p( 2n) $ is also a matrix group
with elements that also may be realized by $2n+2$ matrices.\ \ In
this case a basis ordering, $\{ d p^{i},d q^{i},d \varepsilon ,d
t\}$$=\{d y^{\alpha }, d \varepsilon , d t\}$ is convenient in terms
of which the elements of the group are\ \ 
\begin{equation}
\Gamma ( A,f,v,r) =\left( \begin{array}{lll}
 A & 0 & w \\
 w^{t}\zeta \mbox{}^{\circ} & 1 & r \\
 0 & 0 & 1
\end{array}\right) \in \mathcal{H}\mathcal{S}p( 2n) ,
\end{equation}

\noindent where $A\in \mathcal{S}p( 2n) $ is a $2n$ dimensional
matrix, and $\{w\in \mathbb{R}^{2n},r\in \mathbb{R}\}$ parameterize
the Weyl-Heisenberg normal subgroup.\ \ $w$ may be expanded in terms
of $f,v\in \mathbb{R}^{n}$ as $w=(f,v)\in \mathbb{R}^{2n}$.\ \ $f,v,r$
are force, velocity and power in the classical limit.

This group is the maximal connected group leaving invariant a symplectic
2-form\ \ $\omega =-d e \wedge d t+d p_{i} \wedge d q^{i}$ and degenerate
line element $\gamma \mbox{}^{\circ}=d t^{2}$,\ \ 
\begin{equation}
\varphi {}^{*}\omega =\omega , \varphi {}^{*}\gamma \mbox{}^{\circ}=\gamma
\mbox{}^{\circ}. %
\label{rrr: invariance w dt^2}
\end{equation}

\noindent {\bfseries Theorem:}\ \ {\itshape Let} $\mathbb{P}\simeq
\mathbb{R}^{2n+2}$ {\itshape be extended phase space with a symplectic
2-form }$\omega $ {\itshape and degenerate line element }$\gamma
\mbox{}^{\circ}$.\ \ {\itshape Then, diffeomorphisms} $\varphi$
{\itshape on} $ \mathbb{P}$ {\itshape leaving invariant }$\omega
$ {\itshape and} $\gamma \mbox{}^{\circ}$ {\itshape have a maximum
connected symmetry }
\begin{equation}
\mathcal{H}\mathcal{S}p( 2n) \simeq \mathcal{S}p( 2n+2) \cap \mathcal{I}\mathcal{G}\mathcal{L}(
2n+1,\mathbb{R}) .%
\label{rrr: HSp intersection}
\end{equation}

\noindent {\itshape and locally satisfy Hamilton's equations.} 

The proof of this theorem is given in \cite{Low7}.\ \ It is straightforward
to establish (86) and then locally (85) requires that\ \ $[\frac{\partial
\varphi }{\partial  z}]=\Gamma \in \mathcal{H}\mathcal{S}p( 2n)
$. While not immediately obvious in this form, the resulting first
order differential equations are Hamilton's equations.\ \ 
\subsection{Contraction of the inhomogeneous groups}

\ \ \ On the other hand, the Lorentz group contracts in the limit
$c\rightarrow \infty $\ \ form Special Relativity (SR) to the Euclidean
group $\mathcal{E}( n) $ of Galilean Relativity (GaR) as given in
(15). Note that $\mathcal{E}( n) $ is the subgroup of the affine
group $\mathcal{E}( n) \subset \mathcal{I}\mathcal{G}\mathcal{L}(
n,\mathbb{R}) $ that leaves length invariant. 

\ \ \ The corresponding In\"on\"u-Wigner contraction sequence for
Reciprocal Relativity (RR) is 
\begin{equation}
\begin{array}{lllll}
 \mathrm{RR} &   & \mathrm{SR} &   & \mathrm{GaR} \\
 \mathcal{U}( 1,n)  & \operatorname*{\rightarrow }\limits_{b \rightarrow
\infty } & \mathcal{L}a( 1,n)   & \operatorname*{\rightarrow }\limits_{c
\rightarrow \infty } & \mathcal{H}\mathcal{A}a( n) 
\end{array}.
\end{equation}

\noindent The group $\mathcal{L}a( 1,n) $ as defined in (81) and
the group\ \ 
\begin{equation}
\mathcal{H}\mathcal{A}a( n) \simeq \mathcal{S}\mathcal{O}( n) \otimes
_{s}\left( \mathcal{H}( n) \otimes \mathcal{A}( m) \right) , m=\frac{\left(
n+1\right) \left( n+2\right) }{2}\text{}.
\end{equation}

\noindent The Weyl-Heisenberg group $\mathcal{H}( n) $ are parameterized
by velocity $v$, force $f$ and power $r$ \cite{Low7},\cite{Low8}.\ \ The
abelian subgroup $\mathcal{A}( m) $ is parameterized by the stress
$m^{i,j}$.\ \ \ The abelian group $\mathcal{A}( m) $ is a normal
subgroup and therefore there exits a homomorphism onto the Hamilton
group $\mathcal{H}a( n) $,\ \ 
\begin{equation}
\pi : \mathcal{H}\mathcal{A}a( n) \rightarrow \mathcal{H}a( n) \simeq
\mathcal{S}\mathcal{O}( n) \otimes _{s}\mathcal{H}( n) .
\end{equation}

By Theorem 3 (Appendix A), the complete set of representations of
$\mathcal{H}\mathcal{A}a( n) $ include the representations of its
homomorphic groups. These include $\mathcal{H}a( n) $ and the groups
listed in Appendix A of \cite{Low12}. The stress generators of the
subgroup $\mathcal{A}( m) $ commute will all the other generators
except for the rotation group (spin). Consequently, it appears to
be essentially noninteracting and does not play an essential role
in the classical theory\ \ and therefore we focus in this study
on the Hamilton homomorphic group (See Appendix C for further discussion).\ \ \ 

The Hamilton group is a matrix group with elements
\begin{equation}
\Gamma ( \mathrm{R},f,v,r) =\left( \begin{array}{llll}
 {\mathrm R} & 0 & 0 & f \\
 0 & {\mathrm R} & 0 & v \\
 v & -f & 1 & r \\
 0 & 0 & 0 & 1
\end{array}\right) \in \mathcal{H}a( n) .
\end{equation}

\noindent where $\mathrm{R}\in \mathcal{S}\mathcal{O}( n) $ and
$f,v\in \mathbb{R}^{n}$.\ \ 

The groups each have subgroups that are the expected inertial groups
for transforming between inertial states
\begin{equation}
\begin{array}{llllll}
   & \mathrm{RR} &   & \mathrm{SR} &   & \mathrm{NR} \\
   & \mathcal{U}( 1,n)  & \operatorname*{\rightarrow }\limits_{b
\rightarrow \infty } & \mathcal{L}a( 1,n)   & \operatorname*{\rightarrow
}\limits_{c \rightarrow \infty } & \mathcal{H}\mathcal{A}a( n) 
\\
 \ \  & \cup  &   & \cup  &   & \cup  \\
 \mathrm{Inertial}:  & \mathcal{L}( 1,n)  & \operatorname*{\rightarrow
}\limits_{b \rightarrow \infty } & \mathcal{L}( 1,n)   & \operatorname*{\rightarrow
}\limits_{c \rightarrow \infty } & \mathcal{E}( n) 
\end{array}
\end{equation}
\subsection{Reciprocal relativity classical transformation equations}

In this section, we explore the consequences of the homogeneous
classical (i.e. not quantum) relativity group $\mathcal{U}( 1,n)
$ acting on a basis $\{d z^{\beta }\}$ of the cotangent vector spaces
$T_{z}^{*}\mathbb{P}$ of the extended phase space\ \ $\mathbb{P}\simeq
\mathbb{R}^{2n+2}$.\ \ For $\Gamma ( \Lambda ,\mathrm{M}) \in \mathcal{U}(
1,n) $ that are defined in (77), 
\begin{equation}
d {\widetilde{z}}^{\alpha }={\Gamma ( \Lambda ,\mathrm{M}) }_{\beta
}^{\alpha }d z^{\beta }.%
\label{rrr: U dz transforms}
\end{equation}

\noindent Setting $\{z^{\alpha }\}=\{\frac{x^{a}}{\lambda _{x}},\frac{p^{a}}{\lambda
_{p}}\}$,\ \ the transformations (92) are
\begin{equation}
\begin{array}{l}
 d {\widetilde{x}}^{a}=\Delta ( \Lambda _{b}^{a} d x^{b}-\frac{1}{b}{\mathrm{M}}_{b}^{a}d
p^{b})  \\
 d{\widetilde{p}}^{a}=\Delta ( \Lambda _{b}^{a} d p^{b}+b {\mathrm{M}}_{b}^{a}d
x^{b}) .
\end{array}%
\label{rrr: U x p}
\end{equation}

Consider as an example the case $n=1$. The $\Lambda $ and $\mathrm{M}$
matrices have the solution
\begin{equation}
\Lambda =\gamma \mbox{}^{\circ} \left( \begin{array}{ll}
 1 & \frac{v}{c} \\
 \frac{v}{c} & 1
\end{array}\right) ,\ \ \ \ \ \mathrm{M\mbox{}^{\circ}}=\gamma \mbox{}^{\circ}
\left( \begin{array}{ll}
 \frac{- r}{b c} & \frac{f}{b} \\
 \frac{f}{b} & \frac{r}{b c}
\end{array}\right) ,\ \ \Delta =\gamma /\gamma \mbox{}^{\circ}.%
\label{rrr: 1 d U}
\end{equation}

\noindent where $\gamma ={(1-v^{2}/c^{2}- f^{2}/b^{2}\mathrm{+}r^{2}/b^{2}
c^{2})}^{-1/2}$ and $\gamma \mbox{}^{\circ}={(1-v^{2}/c^{2})}^{-1/2}$.\ \ (Note
that $\Delta =\gamma /\gamma \mbox{}^{\circ}$ holds for general
$n$.)\ \ \ 

Substituting (94) into (93) with the above scaling of the co-ordinates
results in the transformation equations 
\begin{equation}
\begin{array}{l}
 d\widetilde{ t}=\gamma ( d t+\frac{v}{c^{2}}d q+\frac{f}{b^{2}}d p-
\frac{r}{b^{2}c^{2}}d \varepsilon ) , \\
 d\widetilde{ q}=\gamma (  d q+v d t+\frac{r}{b^{2}}d p-\frac{f}{b^{2}}d
\varepsilon ) , \\
 d\widetilde{ p}=\gamma ( d p+ f d t-\frac{r}{c^{2}} d q +\frac{v}{c^{2}}
d \varepsilon ) , \\
 d\widetilde{\varepsilon }=\gamma ( d \varepsilon +v\cdot d p-f\cdot
d q+r d t) .
\end{array}%
\label{rrr: 1 d noninertal tx}
\end{equation}

\noindent In the special case that $f=r=0$, these reduce to the
usual inertial transformation equations of special relativity. 
\begin{equation}
\begin{array}{l}
 d\widetilde{ t}=\gamma \mbox{}^{\circ}( d t+\frac{v}{c^{2}}d q) , \\
 d\widetilde{ q}=\gamma \mbox{}^{\circ}(  d q+v d t) , \\
 d\widetilde{ p}=\gamma \mbox{}^{\circ}( d p +\frac{v}{c^{2}} d \varepsilon
) , \\
 d\widetilde{\varepsilon }=\gamma \mbox{}^{\circ}( d \varepsilon +v
d p) .
\end{array}
\end{equation}

The inertial transformations are block diagonal and the transform
position-time degrees of freedom into themselves and the momentum-energy
degrees of freedom into themselves.\ \ However, the general noninertial
transformations (95) mix all the degrees of freedom. This means
that the position-time (i.e: spacetime) subspace of extended phase
space depends on the noninertial state of the observers.\ \ Now,
the small $f/b$ and $r/b c$ approximation can be determined by taking
the limit $b\rightarrow \infty $ of (95).\ \ \ 
\begin{equation}
\begin{array}{l}
 d\widetilde{ t}=\gamma \mbox{}^{\circ}( d t+\frac{v}{c^{2}}d q) , \\
 d\widetilde{ q}=\gamma \mbox{}^{\circ}(  d q+v d t) , \\
 d\widetilde{ p}=\gamma \mbox{}^{\circ}( d p+ f d t-\frac{r}{c^{2}}
d q +\frac{v}{c^{2}} d \varepsilon ) , \\
 d\widetilde{\varepsilon }=\gamma \mbox{}^{\circ}( d \varepsilon +v\cdot
d p-f\cdot d q+r d t) .
\end{array}
\end{equation}

In this limit, these transformation equations now leave invariant
position-time space (ie spacetime) as an invariant subspace. All
observers, inertial and noninertial, agree on the spacetime subspace
of extended phase space.\ \ This limit regains the notion of an
absolute inertial frame.\ \ \ This gives the illusion of the concept
of an absolute inertial frame.\ \ 

This is analogous to the $c\rightarrow \infty $ limit of special
relativity.\ \ The usual special relativistic spacetime is transformed
into itself by the equations 
\begin{equation}
\begin{array}{l}
 d\widetilde{ t}=\gamma \mbox{}^{\circ}( d t+\frac{v}{c^{2}}d q) , \\
 d\widetilde{ q}=\gamma \mbox{}^{\circ}(  d q+v d t) .
\end{array}
\end{equation}

\noindent It does not have an invariant time subspace and consequently
there is no absolute rest frame. However, in the limit $c\rightarrow
\infty $, time becomes an invariant subspace
\begin{equation}
\begin{array}{l}
 d\widetilde{ t}= d t, \\
 d\widetilde{ q}= d q+v d t,
\end{array}
\end{equation}

\noindent and there is the illusion of an absolute rest frame.

This process may be repeated for the small $f/b$, $v/c$ and $r/c
b$ approximation by taking the limit of both $b,c\rightarrow \infty
$.\ \ For $n=1$, the resulting transformation equations for the
group $\mathcal{H}a( 1) $\ \ are\ \ 
\begin{equation}
\begin{array}{l}
 d\widetilde{ t}=d t, \\
 d\widetilde{ q}= \mathrm{R} d q+v d t, \\
 d\widetilde{ p}=\mathrm{R} d p+ f d t, \\
 d\widetilde{\varepsilon }=d \varepsilon +v\cdot  d p-f\cdot  d q+r
d t.
\end{array}
\end{equation}

The inertial transformation $f=r=0$ are the transformations for
the one dimensional Euclidean group $\mathcal{E}( 1) \subset \mathcal{H}a(
1) $ where $\mathcal{E}( n) $ of homogeneous group of Galilean inertial
relativity as expected, 
\begin{equation}
\begin{array}{l}
 d\widetilde{t}=d t, \\
 d\widetilde{q}=\mathrm{R} d q + v d t, \\
 d\widetilde{p}=\mathrm{R} d p,  \\
 d \widetilde{\varepsilon } = d \varepsilon  + v d p. 
\end{array}%
\label{mo: nonrelativistic inertial}
\end{equation}

These expression generalize to general $n$ and the group contraction
is determined from the In\"on\"u-Wigner contraction of the Lie algebra
as given in the Appendix B.\ \ \ 

We note also that in {\itshape four} notation,\ \ the matrix elements
of the group $\mathcal{L}a( 1,n) $ are given in (83). The corresponding
transformation equations are
\begin{equation}
\begin{array}{l}
 d {\widetilde{x}}^{a}=\Lambda _{b}^{a} d x^{b} \\
 d{\widetilde{p}}^{a}=\Lambda _{b}^{a} d p^{b}+ {\mathrm{M\mbox{}^{\circ}}}_{b}^{a}d
x^{b}.
\end{array}
\end{equation}

\noindent Again, the inertial case is 
\begin{equation}
\begin{array}{l}
 d {\widetilde{x}}^{a}=\Lambda _{b}^{a} d x^{b}, \\
 d{\widetilde{p}}^{a}=\Lambda _{b}^{a} d p^{b},
\end{array}
\end{equation}

\noindent with
\begin{equation}
\Gamma \mbox{}^{\circ}( \Lambda ,0) =\left( \begin{array}{ll}
 \Lambda  & 0 \\
 0 & \Lambda 
\end{array}\right) \in \mathcal{L}( 1,n) .
\end{equation}

In these equations, $\mathrm{M}$ is physically interpreted as the
power-force stress tensor that is symmetric in the sense that ${\mathrm{M}}_{a}^{b}=\eta
_{a,c}\eta ^{b,d}{\mathrm{M}}_{d}^{c}$ and therefore has $m=\frac{(n+1)(n+2)}{2}$
independent components. This is the expected transform to a noninertial
frame in special relativistic quantum mechanics.\ \ In this limit,
spacetime is again an invariant subspace of the extended phase space
and consequently the inertial frame is again absolute, independent
of the noninertial state.\ \ 
\section{Reciprocally relativistic quantum mechanics}\label{rrr:
Chapter RR QM}

In the introductory Section 1.2, we reviewed how special relativistic
quantum mechanics is given by the projective representations of
the inhomogeneous Lorentz group $\mathcal{I}\mathcal{L}( 1,n) $.\ \ These
representations are equivalent to the unitary representations of
the Poincar\'e group that is the central extension of the inhomogeneous
Lorentz group. This is also the case in Galilean relativistic quantum
mechanics. It is defined by the projective representations of the
inhomogeneous Euclidean group $\mathcal{I}\mathcal{E}( n) $ that
are equivalent to the unitary representations of the cover of the
Galilei group that is the central extension of $\mathcal{I}\mathcal{E}(
n) $.

This is true also for the symmetry $\mathcal{U}( 1,n) $ on extended
phase space.\ \ The reciprocal relativistic quantum theory is given
in terms of the projective representations of the inhomogeneous
group unitary group
\begin{equation}
 \mathcal{I}\mathcal{U}( 1,n) \simeq \mathcal{U}( 1,n) \otimes \mathcal{A}(
2n+2) .
\end{equation}

\noindent \ \ Furthermore, this is true also for the limiting forms
of the inhomogeneous groups\ \ \ 
\begin{equation}
\begin{array}{lllll}
 \mathrm{RR} &   & \mathrm{SR} &   & \mathrm{GaR} \\
 \mathcal{I}\mathcal{U}( 1,n)  & \operatorname*{\rightarrow }\limits_{b
\rightarrow \infty } & \mathcal{I}\mathcal{L}a( 1,n)   & \operatorname*{\rightarrow
}\limits_{c \rightarrow \infty } & \mathcal{I}\mathcal{H}\mathcal{A}a(
n) 
\end{array}
\end{equation}

\noindent where 
\begin{gather}
\mathcal{I}\mathcal{L}a( 1,n) =\mathcal{L}a( 1,n) \otimes \mathcal{A}\left(
2n+2\right) 
\\\mathcal{I}\mathcal{H}\mathcal{A}a( n) =\mathcal{H}\mathcal{A}a(
n) \otimes \mathcal{A}\left( 2n+2\right) 
\end{gather}

The projective representations are computed using Theorem 2 (Appendix
A) from the unitary representations of their central extension.\ \ In
each of these cases, the central extension turns the abelian normal
subgroup $\mathcal{A}( 2n+2) $ into a Weyl-Heisenberg group $\mathcal{H}(
n+1) $. The hermitian representation of the algebra of this Weyl-Heisenberg
subgroup is precisely the Heisenberg commutation relations for position
and momentum and energy and time.\ \ The unitary irreducible representations
are computed using the Mackey theorems for the case where the normal
subgroup is the nonabelian Weyl-Heisenberg group. 
\subsection{The quantum symmetry: Relativity implications of the
quantum phase}

The projective representations of the inhomogeneous unitary group
$\mathcal{I}\mathcal{U}( 1,n) $ are equivalent to the unitary representations
of its central extension
\begin{equation}
\begin{array}{rl}
 \widecheck{\mathcal{I}\mathcal{U}}( 1, n)  & \simeq \overline{\mathcal{U}}(
1, n) \otimes _{s}\mathcal{H}( n+1) \subset {\mathcal{A}ut}_{\mathcal{H}(
n+1) } \\
  & =\mathcal{D}\otimes _{s}\mathcal{S}\mathcal{U}( 1, n) \otimes
_{s}\mathcal{H}( n+1) 
\end{array}
\end{equation}

\noindent where $\overline{\mathcal{U}}( 1) \simeq \mathcal{D}$.\ \ The
central extension has a single algebraic generator that results
in the abelian subgroup $\mathcal{A}( 2n+2) $ becoming the Weyl-Heisenberg
group $\mathcal{H}( n+1) $. The brackets of the hermitian representation
of the Lie algebra of this Weyl-Heisenberg subgroup are precisely
the Heisenberg commutation relations. The central extension of the
inhomogeneous unitary group $\widecheck{\mathcal{I}\mathcal{U}}( 1,
n) $ is a subgroup of the automorphism group of the Weyl-Heisenberg
group (40).\ \ It is convenient to define a group that we refer
to as the quaplectic group as the algebraic part of the central
extension,\ \ \ 
\begin{equation}
\mathcal{Q}( 1,n) \simeq \mathcal{U}( 1, n) \otimes _{s}\mathcal{H}(
n+1) 
\end{equation}

\noindent so that $\widecheck{\mathcal{I}\mathcal{U}}( 1, n) \simeq
\overline{\mathcal{Q}}( 1,n) $.

 The computation of the unitary irreducible representations of $\widecheck{\mathcal{I}\mathcal{U}}(
1, n) $ requires the full power of the nonabelian Mackey theorems.\ \ These
unitary irreducible representations have been computed in \cite{Wolf},\cite{Low4},\cite{Low6}.\ \ \ 

The resulting Hilbert space of states for the faithful representation
is ${\text{\boldmath $L$}}^{2}( \mathbb{R}^{n+1},\text{\boldmath
$\mathrm{H}$}) $. $\text{\boldmath $\mathrm{H}$}$ is the Hilbert
space for the ordinary unitary representations of $\overline{\mathcal{U}}(
1,n) $ that is a countably infinite complex vector space $\text{\boldmath
$\mathrm{H}$}\simeq \mathbb{V}^{\infty }$ \cite{Ottoson}.\ \ The
wave functions are functions of commuting subset of $\mathcal{H}(
n+1) $ and not the full phase space.\ \ One case is $\psi ( t,q)
$ with $\{t,q\}\in \mathbb{R}^{n+1}$ parameterizing an abelian subgroup
$\mathcal{A}( n+1) $ of the Weyl-Heisenberg group $\mathcal{H}(
n+1) $.\ \ Other cases are $\psi ( e,p) $, $\psi ( t,p) $ or $\psi
( e,q) $.\ \ 

The Hilbert space is over $\mathbb{R}^{n+1}$ and not the $n$ dimensional
mass shell hypersurface as in the inhomogeneous Lorentz group.\ \ The
`mass shell' is statistically determined by the wave function, it
is not a hypersurface. Furthermore, note that the special relativistic
concept of mass is not invariant in this theory. 
\subsection{The $b\rightarrow \infty $ limit}

Projective representations of $\mathcal{I}\mathcal{L}a( 1,n) $ are
unitary representations of $\widecheck{\mathcal{I}\mathcal{L}a}( 1,n)
=\overline{\mathcal{L}a}( 1,n) \otimes _{s}\mathcal{H}( n+1) $ \cite{Low9}.\ \ Again,
the hermitian representation of the algebra of the Weyl-Heisenberg
subgroup are the Heisenberg commutation relations. 

From Theorem 3 (Appendix A),\ \ the complete set of unitary irreducible
representations includes the degenerate representations that are
faithful representations of homomorphic groups. These groups are
as follows. First,\ \ 
\begin{equation}
\begin{array}{ll}
 \pi :\mathcal{I}\mathcal{L}a( 1,n) \rightarrow \mathcal{I}\mathcal{L}(
1,n)  & \ker ( \pi ) \simeq \mathcal{A}( m) \otimes \mathcal{A}(
n+1) 
\end{array}
\end{equation}

\noindent where $m=\frac{(n+1)(n+2)}{2}$. The normal subgroup $\mathcal{A}(
m) \otimes \mathcal{A}(n+1)$ has the basis of generators $\{M \mbox{}^{\circ}_{a,b},X_{a}\}$
as given in Appendix B (126). The group $\mathcal{I}\mathcal{L}(
1,n) $ has homomorphic groups $\mathcal{L}( 1,n) $ and of course
the trivial group.\ \ \ In addition, the group $\mathcal{I}\mathcal{L}a(
1,n) $ is homomorphic to its homogeneous group $\mathcal{L}a( 1,n)
$. 

In standard special relativity, in considering the projective representations
of the inhomogeneous Lorentz group,\ \ we do not consider the representations
of the homogeneous\ \ group to be physical and disregard them in
the physical discussion \cite{Weinberg1}.\ \ This is true also in
this case for the homogeneous degenerate group $\mathcal{L}a( 1,n)
$. 

The remaining homomorphic group $\mathcal{I}\mathcal{L}( 1,n) $
is precisely the inhomogeneous Lorentz group of special relativity.\ \ Therefore,
the projective representations of $\mathcal{I}\mathcal{L}a( 1,n)
$ include the expected inertial representations of the inhomogeneous
Lorentz group in this regime of interactions that are small relative
to $b$. 

There are however other physical noninertial states in the faithful
projective representations of $\mathcal{I}\mathcal{L}a( 1,n) $ that
embody energy.\ \ Given that we only appear to see a subset of the
mass an energy in the universe, this may be interesting to investigate
further.\ \ \ Some additional insight to this is provided in the
following section. 
\subsection{The $b,c\rightarrow \infty $ limit: the quantum Hamilton
group}

The quantum Hamilton group corresponds to the limit where both $b,c\rightarrow
\infty $\ \ and so forces are small relative to $b$ and velocities
are small relative to $c$.\ \ The quantum theory is the projective
representations of the inhomogeneous Hamilton group\footnote{The
most general case is the $\mathcal{I}\mathcal{H}\mathcal{A}a( n)
$ group discussed in Appendix C.} [Low12],[Low8]
\begin{equation}
\mathcal{I}\mathcal{H}a( n) \simeq \mathcal{H}a( n) \otimes _{s}\mathcal{A}(
2n+2) .
\end{equation}

\noindent These are equivalent to the unitary representation of
the central extension
\begin{equation}
\widecheck{\mathcal{I}\mathcal{H}a}( n) \simeq \overline{\mathcal{H}a}(
n) \otimes _{s}\left( \mathcal{H}( n+1) \otimes \mathcal{A}( 2)
\right) .
\end{equation}

There are three central elements in the algebraic central extension
\cite{Low8}. First there is the generator $I$ that is the central
generator of the Weyl-Heisenberg group $\mathcal{H}( n+1) $.\ \ Second,
there is a central extension $M$ that is mass and is one of the
two generators of the $\mathcal{A}( 2) $ subgroup.\ \ The Galilei
group $\overline{\mathcal{G}a}( n) $ is the inertial subgroup of\ \ $\widecheck{\mathcal{I}\mathcal{H}a}(
n) $ and this is precisely the same mass generator (17) for\ \ $\overline{\mathcal{G}a}(
n) \subset \widecheck{\mathcal{I}\mathcal{H}a}( n) $.\ \ 

Finally there is a third central generator of the central extension
denoted $A$ that is the second generator of the $\mathcal{A}( 2)
$ subgroup.\ \ This generator is a new prediction of the theory.
It has dimensions of the reciprocal of tension (in MKS units, m/Newton).\ \ \ 

Given how fundamental the central generators $I$ and $M$ are to
physics, we expect $A$ to be fundamental also. $A$ interacts through
a non-inertial generalization to usual `nonrelativistic' spin \cite{Low8}.
It embodies energy $A b^{2}$ just as mass embodies energy $M c^{2}$.
In the full reciprocal relativistic theory they {\itshape combine}
into a generalized power-force stress `tensor' realized by the generators
$M_{a,b}$ (125).\ \ This is a definitive prediction of the theory;
the central generators $A$ is a `residue' of the full theory in
its contraction to the `nonrelativistic' domain. It should be possible
to detect $A$ using the above mentioned noninertial spin interaction.\ \ 
\section{Concluding remarks}

This paper has explored the relativity implications of the quantum
phase. The quantum phase required us to consider projective representations
of the symmetry groups that are equivalent to the unitary representations
of their central extensions. This establishes the equivalence of
the unitary representations of the Weyl-Heisenberg group with a
particular projective representation of the abelian translation
group on extended phase space.\ \ This leads to the projective representations
of the conformally scaled inhomogeneous symplectic group as the
largest representation in which the Heisenberg commutation relations
are valid at all states in the Hilbert space under the action of
the unitary representations of the central extension of the group.\ \ 

Requiring relativistic concepts of invariant time through an orthogonal
line element leads to the Born metric on extended phase space with
the fundamental scale constant $b$.\ \ The resulting symmetry is
the inhomogeneous unitary group $\mathcal{I}\mathcal{U}( 1,n) $.\ \ This
symmetry describes the transformation between noninertial, as well
as inertial, states. Spacetime in this theory is relative to noninertial
states and there is neither an absolute inertial nor rest frame.
The limit $b\rightarrow \infty $ results in spacetime becoming an
invariant subspace with the associated appearance of an absolute
inertial frame.\ \ The full classical limit with $c,b\rightarrow
0$ yields a new derivation of Hamilton' s equations. These transformation
equations have absolute Newtonian time with the appearance of an
absolute inertial and rest frame.\ \ \ 

The concepts are presented as global symmetries for an extended
phases space $\mathbb{P}\simeq \mathbb{R}^{2n+2}$.\ \ The symmetries
need to be made local and applies to more general manifolds to provide
a generalization of relativity where the local symmetry is $\mathcal{U}(
1,n) $ rather than $\mathcal{L}( 1,n) $.\ \ This has been explored
in a series of papers by Mantz et al \cite{Mantz-1},\cite{Mantz-2},\cite{burgers}

The noninertial quantum theory is the projective representation
of the inhomogeneous unitary group. (Understanding noninertial quantum
mechanics,\ \ is a prelude to understanding quantum gravity.) The
limits yield special relativistic quantum mechanics in the limit
$b\rightarrow \infty $ and quantum Hamiltonian mechanics in the
limit $c,b\rightarrow \infty $.\ \ This noninertial quantum symmetry
resulted from the relativistic implications of the quantum phase.

I would like to thank Peter Jarvis, Stuart Morgan and Rutwig Campoamor-Stursberg
for discussions during the course of this work.\ \ I would like
to thank C\'ecile DeWitt-Morette for pointing out the Dirac quote
and for her encouragement of this work. 
\section{Appendix A: Basic theorems of projective representations}

The following set of theorems, together with the Mackey theorems
\cite{mackey},\cite{Low12} enables the projective representations
for a general class of Lie groups to be calculated.
\begin{theorem}

{\bfseries (Wigner, Weinberg)} Any projective representation of
a Lie symmetry group $\mathcal{G}$ on a separable Hilbert space
is equivalent to a representation that is either linear and unitary
or anti-linear and anti-unitary. Furthermore, if $\mathcal{G}$ is
connected, the projective representations are equivalent to a representation
that is linear and unitary \cite{wignerb},\cite{Weinberg1}.\label{PH:
theorem: Wigner unitary projective}
\end{theorem}

This is the generalization of the well known theorem that the ordinary
representation of any compact group is equivalent to a representation
that is unitary.\ \ For a projective representation, the phase degrees
of freedom of the central extension enables the equivalent linear
unitary or antilinear antiunitary representation to be constructed
for this much more general class of Lie groups that admit representations
on separable Hilbert spaces.\ \ (A proof of the theorem is given
in Appendix A of Chapter 2 of \cite{Weinberg1}).\ \ The groups that
this theorem applies to include all groups that are mentioned in
this paper.
\begin{theorem}

{\bfseries (Bargmann, Mackey)} The projective representations of
a connected Lie group $ \mathcal{G}$ are equivalent to the ordinary
unitary representations of its central extension $\widecheck{\mathcal{G}}$
\cite{bargmann},\cite{mackey2}.\label{PH: theorem: proj rep is unitary
CE}
\end{theorem}

Theorem 1 states that are all projective representations are equivalent
to a projective representation that is unitary. A phase is the unitary
representation of a central abelian subgroup. Therefore, the maximal
representation is given in terms of the central extension of the
group. For simply connected groups, this definition is equivalent
to the formulation of a projective representation as an ordinary
unitary representation that is defined {\itshape up to a phase},\ \ $\varrho
( \gamma ^{\prime }) \varrho ( \gamma ) =e^{i \theta }\varrho (
\gamma ^{\prime }\gamma ) $\cite{bargmann},\cite{mackey2}.\ \ (Note
that all centrally extended groups are simply connected.)\ \ \ 
\begin{theorem}

Let $\mathcal{G}$,$\mathcal{H}$ be Lie groups and $\pi :\mathcal{G}\rightarrow
\mathcal{H}$ be a homomorphism. Then, for every unitary representation
$\widetilde{\varrho }$ of $\mathcal{H}$ there exists a degenerate unitary
representation $\varrho $ of $\mathcal{G}$ defined by $\varrho =\widetilde{\varrho
}\circ \pi $. Conversely, for every degenerate unitary representation
of a Lie group $\mathcal{G}$ there exists a Lie subgroup $\mathcal{H}$
and a homomorphism $\pi :\mathcal{G}\rightarrow \mathcal{H}$ where
$\ker ( \pi ) \neq \text{\boldmath $e$}$ such that $\varrho =\widetilde{\varrho
}\circ \pi $\ \ where $\widetilde{\varrho }$ is a unitary representation
of $\mathcal{H}$.\label{PH: theorem: degenerate reps}
\end{theorem}

Noting that a representation is a homomorphism, This theorem follows
straightforwardly from the properties of homomorphisms. As a consequence,
the set of degenerate representations of a group is characterized
by its set of normal subgroups. A {\itshape faithful} representation
is the case that the representation is an isomorphism.
\begin{theorem}

{\bfseries (Levi) }Any simply connected\ \ Lie group is equivalent
to the semidirect product of a semisimple group and a maximal solvable
normal subgroup \cite{barut}.\label{PH: theorem: Levi}
\end{theorem}

As the central extension of any connected group is simply connected,
the problem of computing the projective representations of a group
always can be reduced to computing the unitary irreducible representations
of a semidirect product group with a semisimple homogeneous group
and a solvable normal subgroup.\ \ The unitary irreducible representations
of the semisimple groups are known and the solvable groups that
we are interested in turn out to be the semidirect product of abelian
groups.
\begin{theorem}

Any semidirect product group $\mathcal{G}\simeq \mathcal{K}\otimes
_{s}\mathcal{N}$ is a subgroup of a group homomorphic to the group
of automorphisms of the normal subgroup, $\mathcal{G}\subset {\mathcal{A}ut}_{\mathcal{N}}$
\cite{barut}.\label{PH: theorem: automorphisms semid-direct}
\end{theorem}

This theorem places constraints on the admissible semidirect product
groups that have a given normal subgroup.\ \ For example, the automorphism
group of the abelian group is
\begin{equation}
 {\mathcal{A}ut}_{\mathcal{A}( m) }\simeq \mathbb{Z}_{2}\otimes
_{s}\mathcal{D}\otimes _{s}\overline{\mathcal{G}\mathcal{L}}( m,\mathbb{R})
\otimes _{s}\mathcal{A}( m)  , \mathcal{D}\simeq \left( \mathbb{R}^{+},\times
\right) ,%
\label{PH: Abelain automorphism group}
\end{equation}

\noindent whereas the automorphism group of the Weyl-Heisenberg
group $\mathcal{H}( m) $ is \cite{folland},\cite{Low8}
\begin{equation}
 {\mathcal{A}ut}_{\mathcal{H}( m) }\simeq \mathbb{Z}_{2}\otimes
_{s}\mathcal{D}\otimes _{s}\overline{\mathcal{S}p}( 2m) \otimes
_{s}\mathcal{H}( m) .%
\label{PH: Weyl-Heisenberg automorphism group}
\end{equation}

This means that there does not exist a semidirect product of the
form ``$\mathcal{S}\mathcal{O}( 2m) \otimes _{s}\mathcal{H}( m)
$ `` as $\mathcal{S}\mathcal{O}( 2m) $ is not a subgroup of $\mathcal{S}p(
2m) $. On the other hand, the semidirect product $\mathcal{H}a(
n) =\mathcal{S}\mathcal{O}( m) \otimes _{s}\mathcal{H}( m) $ is
admissible as $\mathcal{S}\mathcal{O}( m) \subset \mathcal{S}p(
2m) $.$\text{}$
\section{Appendix B: Algebra of $\mathcal{Q}( 1,n) $ and In\"on\"u-Wigner
contractions }

This appendix defines the algebra of the quaplectic group and its
contraction in the $b\rightarrow \infty $ and $c\rightarrow \infty
$.\ \ \ A general element of the algebra may be written in terms
of dimensionless parameters as
\begin{equation}
Z=\alpha ^{i,j}J_{i,j}+{\widetilde{\beta }}^{i}K_{i}+{\widetilde{\gamma
}}^{i}N_{i}+ \widetilde{r} R+{\widetilde{m}}^{i,j}M_{i,j}+{\widetilde{q}}^{i}P_{i}+{\widetilde{p}}^{i}Q_{i}+\widetilde{t}
E+\widetilde{\varepsilon } T+\iota  I.
\end{equation}

\noindent where $i,j,...=1,...,n.$ Scaling the dimensionless parameters
with appropriate physical dimensions,\ \ 
\begin{equation}
\begin{array}{lllll}
 \alpha ^{i,j}\mapsto \alpha ^{i,j} & {\widetilde{\beta }}^{i}\mapsto
\beta ^{i}/c & {\widetilde{\gamma }}^{i}\mapsto \gamma ^{i}/b & {\widetilde{r}}^{i}\mapsto
r^{i}/b c & {\widetilde{m}}^{i,j}\mapsto m^{i,j}/b c \\
 {\widetilde{p}}^{i}\mapsto p^{i}/\lambda _{p} & {\widetilde{q}}^{i}\mapsto
q^{i}/\lambda _{q} & \widetilde{t}\mapsto t/\lambda _{t} & \widetilde{\varepsilon
}\mapsto \varepsilon /\lambda _{\varepsilon } & \iota \mapsto \iota
.
\end{array}
\end{equation}

\noindent requires the corresponding scaling of the generators,
\begin{equation}
\begin{array}{lllll}
 J_{i,j}\mapsto J_{i,j} & K_{i}\mapsto c K_{i} & N_{i}\mapsto b
N_{i} & R\mapsto b c R & M_{i,j}\mapsto b c M_{i,j} \\
 P_{i}\mapsto \lambda _{p} P_{i} & Q_{i}\mapsto \lambda _{q} Q_{i}
& T\mapsto \lambda _{t} T & E\mapsto \lambda _{\varepsilon } E &
I\mapsto I.
\end{array}
\end{equation}

The nonzero algebra commutation relations are computed to be
\begin{gather*}
\begin{array}{l}
 \left[ J_{i,j},J_{k,l}\right] =-J_{j,l} \delta _{i,k}+J_{j,k} \delta
_{i,l}+J_{i,l} \delta _{j,k}-J_{i,k} \delta _{j,l}, \\
 \left[ J_{i,j},M_{k,l}\right] =-M_{j,l} \delta _{i,k}-M_{j,k} \delta
_{i,l}+M_{i,l} \delta _{j,k}+M_{i,k} \delta _{j,l},
\end{array}
\\\begin{array}{lll}
 \left[ J_{i,j},K_{k}\right] =-K_{j} \delta _{i,k}+K_{i} \delta
_{j,k}, & \left[ K_{i},K_{k}\right] =\frac{1}{c^{2}}J_{i,k}, & \left[
K_{i},R\right] =-\frac{2 }{c^{2}}N_{i}, \\
 \left[ J_{i,j},N_{k}\right] =-N_{j} \delta _{i,k}+N_{i} \delta
_{j,k}, & \left[ N_{i},N_{k}\right] =\frac{1}{b^{2}}J_{i,k}, & \left[
N_{i},R\right] =\frac{2 }{b^{2}}K_{i}, \\
 \left[ K_{i},N_{k}\right] =-M_{i,k}- \delta _{i,k}R, &   &  
\end{array}
\\\begin{array}{ll}
 \left[ K_{i},M_{k,l}\right] =-\frac{1}{c^{2}}\left( N_{l} \delta
_{i,k}+N_{k} \delta _{i,l}\right) , & \left[ N_{i},M_{k,l}\right]
=\frac{1}{b^{2}}\left( K_{l} \delta _{i,k}+K_{k} \delta _{i,l}\right)
, \\
 \left[ J_{i,j},Q_{k}\right] =-Q_{j} \delta _{i,k}+Q_{i} \delta
_{j,k}, & \left[ J_{i,j},P_{k}\right] =-P_{j} \delta _{i,k}+P_{i}
\delta _{j,k}, \\
 \left[ K_{i},M_{k,l}\right] =-\frac{1}{c^{2}}\left( N_{l} \delta
_{i,k}+N_{k} \delta _{i,l}\right) , & \left[ N_{i},M_{k,l}\right]
=\frac{1}{b^{2}}\left( K_{l} \delta _{i,k}+K_{k} \delta _{i,l}\right)
, \\
 \left[ M_{i,j},Q_{k}\right] =\frac{1}{b^{2}}\left(  \delta _{i,k}P_{j}+
\delta _{j,k}P_{i}\right) , & \left[ M_{i,j},P_{k}\right] =-\frac{1}{c^{2}}\left(
\delta _{i,k}Q_{j}+ \delta _{j,k}Q_{i}\right) ,
\end{array}
\\\begin{array}{lll}
 \left[ K_{i},Q_{k}\right] =- \delta _{i,k}T, & \left[ K_{i},P_{k}\right]
=-\frac{ 1}{c^{2}} \delta _{i,k}E, & \left[ K_{i},T\right] =-\frac{1}{c^{2}}Q_{i},
\\
 \left[ N_{i},P_{k}\right] =- \delta _{i,k}T, & \left[ N_{i},Q_{k}\right]
=\frac{1}{b^{2}} \delta _{i,k}E, & \left[ N_{i},T\right] =-\frac{1}{b^{2}}P_{i},
\\
 \left[ K_{i},E\right] =-P_{i}, & \left[ N_{i},E\right] =Q_{i},
& \left[ R,T\right] =-\frac{2 }{b^{2} c^{2}}E, \\
 \left[ R,E\right] =2 T, & \left[ Q_{i},P_{k}\right] =- \hbar  \delta
_{i,k}I, & \left[ T,E\right] = \hbar  I.
\end{array}
\end{gather*}

Note that there are four different inhomogeneous Lorentz algebras
with generators\ \ $\{J_{i,j},K_{i},P_{i},E\}$, $\{J_{i,j},K_{i},Q_{i},T\}$,
$\{J_{i,j},N_{i},P_{i},T\}$ and $\{J_{i,j},N_{i},Q_{i},E\}$ corresponding
to four different connected Lorentz subgroups $\mathcal{L}( 1,n)
\subset \mathcal{U}( 1,n) $.\ \ 

The limits are computed straightforwardly as In\"on\"u-Wigner contractions.\ \ For
example the $b,c\rightarrow \infty $ limit is 
\begin{gather*}
\begin{array}{l}
 \left[ J_{i,j},J_{k,l}\right] =-J_{j,l} \delta _{i,k}+J_{j,k} \delta
_{i,l}+J_{i,l} \delta _{j,k}-J_{i,k} \delta _{j,l}, \\
 \left[ J_{i,j},M_{k,l}\right] =-M_{j,l} \delta _{i,k}-M_{j,k} \delta
_{i,l}+M_{i,l} \delta _{j,k}+M_{i,k} \delta _{j,l},
\end{array}
\\\begin{array}{l}
 \left[ J_{i,j},K_{k}\right] =-K_{j} \delta _{i,k}+K_{i} \delta
_{j,k}, \\
 \left[ J_{i,j},N_{k}\right] =-N_{j} \delta _{i,k}+N_{i} \delta
_{j,k}, \\
 \left[ K_{i},N_{k}\right] =-M_{i,k}- \delta _{i,k}R,
\end{array}
\\\begin{array}{ll}
 \left[ J_{i,j},Q_{k}\right] =-Q_{j} \delta _{i,k}+Q_{i} \delta
_{j,k}, & \left[ J_{i,j},P_{k}\right] =-P_{j} \delta _{i,k}+P_{i}
\delta _{j,k},
\end{array}%
\label{rrr: Hamiltan A algebra}
\\\begin{array}{lll}
 \left[ K_{i},Q_{k}\right] =- \delta _{i,k}T, &  \left[ N_{i},P_{k}\right]
=- \delta _{i,k}T, &   \\
 \left[ K_{i},E\right] =-P_{i}, & \left[ N_{i},E\right] =Q_{i},
&   \\
 \left[ R,E\right] =2 T, & \left[ Q_{i},P_{k}\right] =- \hbar  \delta
_{i,k}I, & \left[ T,E\right] = \hbar  I.
\end{array}
\end{gather*}

This may be used to establish the contraction diagram
\begin{equation}
\begin{array}{lllllll}
   & \nearrow  & \operatorname*{\rightarrow }\limits_{b\rightarrow
\infty } & \mathcal{L}a( 1,n)  & \operatorname*{\rightarrow }\limits_{c\rightarrow
\infty } &  \searrow  &   \\
  \mathcal{U}( 1,n)  &   &   &   &   & \ \  & \mathcal{H}\mathcal{A}a(
n)   \\
   &  \searrow  & \operatorname*{\rightarrow }\limits_{c\rightarrow
\infty } & \widetilde{\mathcal{L}a}( 1,n)  & \operatorname*{\rightarrow
}\limits_{b\rightarrow \infty } &  \nearrow  &  
\end{array}
\end{equation}

\noindent $\mathcal{L}a( 1, n) \simeq \widetilde{\mathcal{L}a}( 1,n)
$ as a group.\ \ However the generators of the Lorentz subgroup
of $\mathcal{L}a( 1, n) $ are $\{J_{i,j},K_{i}\}$ whereas for $\widetilde{\mathcal{L}a}(
1,n) $, they are $\{J_{i,j},N_{i}\}$.

A corresponding {\itshape four} notation for the generators may
also be defined.\ \ Setting $c=1$, a general element of the algebra
can be written as
\begin{equation}
Z= \alpha ^{a,c}L_{a,c}+\varphi ^{a,c}M_{a,c}+p^{a}X_{a}+x^{a}P_{a}+\iota
I
\end{equation}

\noindent where $a,b, ... =0,....,n$. The scaling of the parameters
\begin{equation}
\begin{array}{lll}
 \alpha ^{a,b}\mapsto \alpha ^{a,b} & {\widetilde{\varphi }}^{a,b}\mapsto
\varphi ^{a,b}/b & \iota \mapsto \iota  \\
 {\widetilde{p}}^{a}\mapsto p^{a}/\lambda _{p} & {\widetilde{x}}^{a}\mapsto
x^{a}/\lambda _{x}. &  
\end{array}
\end{equation}

\noindent requires the corresponding scaling of the generators 
\begin{equation}
\begin{array}{lll}
 L_{a,b}\mapsto L_{a,b} & M_{a,b}\mapsto a M_{a,b} & I\mapsto I
\\
 P_{a}\mapsto \lambda _{p} P_{a} & X_{a}\mapsto \lambda _{x} X_{a}.
&  
\end{array}
\end{equation}

Then the commutation relations are
\begin{gather}
\begin{array}{l}
 \left[ L_{a,e},L_{c,d}\right] =-L_{e,d} \eta _{a,c}+L_{e,c} \eta
_{a,d}+L_{a,d} \eta _{e,c}-L_{a,c} \eta _{e,d}, \\
 \left[ L_{a,e},M_{c,d}\right] =-M_{e,d} \eta _{a,c}-M_{e,c} \eta
_{a,d}+M_{a,d} \eta _{e,c}+M_{a,c} \eta _{e,d}, \\
 \left. \left[ M_{a,e},M_{c,d}\right] =\frac{1}{b^{2}}\left( -L_{e,d}
\eta _{a,c}-L_{e,c} \eta _{a,d}-L_{a,d} \eta _{e,c}-L_{a,c} \eta
_{e,d}\right. \right) ,
\end{array}%
\label{rrr: iu four notation}
\\\begin{array}{ll}
 \left[ L_{a,d},X_{c}\right]  = -X_{d} \eta _{a,c}+X_{a} \eta _{d,c},
& \left[ L_{a,d},P_{c}\right]  = -P_{d} \eta _{a,c}+P_{a} \eta _{d,c}
, \\
 \left. \left[ M_{a,d},X_{c}\right]  = -\frac{1}{b^{2}}\left( P_{d}
\eta _{a,c}+P_{a} \eta _{d,c}\right. \right)  & \left[  M_{a,d},P_{c}\right]
= X_{d} \eta _{a,c}+X_{a} \eta _{d,c}, \\
 \left[ X_{a},P_{c}\right]  = I \eta _{a,c}. &  
\end{array}
\end{gather}

In the limit $b\rightarrow \infty $, this contracts to the nonzero
commutators
\begin{gather}
\begin{array}{l}
 \left[ L_{a,e},L_{c,d}\right] =-L_{e,d} \eta _{a,c}+L_{e,c} \eta
_{a,d}+L_{a,d} \eta _{e,c}-L_{a,c} \eta _{e,d}, \\
 \left[ L_{a,e},M \mbox{}^{\circ}_{c,d}\right] =-M \mbox{}^{\circ}_{e,d}
\eta _{a,c}-M \mbox{}^{\circ}_{e,c} \eta _{a,d}+M \mbox{}^{\circ}_{a,d}
\eta _{e,c}+M \mbox{}^{\circ}_{a,c} \eta _{e,d},
\end{array}%
\label{rrr: four la generators}
\\\begin{array}{ll}
 \left[ L_{a,d},X_{c}\right]  = -X_{d} \eta _{a,c}+X_{a} \eta _{d,c},
& \left[ L_{a,d},P_{c}\right]  = -P_{d} \eta _{a,c}+P_{a} \eta _{d,c}
, \\
 \left[ X_{a},P_{c}\right]  = I \eta _{a,c}, & \left[  M \mbox{}^{\circ}_{a,d},P_{c}\right]
= X_{d} \eta _{a,c}+X_{a} \eta _{d,c}. \\
   &  
\end{array}
\end{gather}

\noindent This is the algebra of the $\mathcal{L}a( 1,n) $ group.\ \ A
dual four notation exists for $\widetilde{\mathcal{L}a}( 1,n) $.
\section{Appendix C: The group $\mathcal{I}\mathcal{H}\mathcal{A}a(
n) $ }

The full quantum theory is the projective representations of the
inhomogeneous group $\mathcal{I}\mathcal{H}\mathcal{A}a( n) $,\ \ 
\begin{equation}
\mathcal{I}\mathcal{H}\mathcal{A}a( n) \simeq \mathcal{S}\mathcal{O}(
n) \otimes _{s}\left( \mathcal{H}( n) \otimes \mathcal{A}( m) \right)
\otimes _{s}\mathcal{A}( 2n+2) ,
\end{equation}

\noindent where $m=\frac{n( n+1) }{2}$. $\mathcal{A}( m) $ is generated
by the generators $M_{i,j}$ given in Appendix B (120) above. It
can be shown that $\mathcal{A}( m) $ is a normal subgroup and therefore
there is a homomorphism
\begin{equation}
\pi :\mathcal{I}\mathcal{H}\mathcal{A}a( n) \rightarrow \mathcal{I}\mathcal{H}a(
n) ,\ \ \ker  \left( \pi \right) \simeq \mathcal{A}( m\text{}) .%
\label{rrr: IHAa IHa homomorphism}
\end{equation}

Again, from Theorem 3 (Appendix A),\ \ the complete set of projective
representations of $\mathcal{I}\mathcal{H}\mathcal{A}a( n)  $ includes
the projective representations of the homomorphic groups. These
are (128)\ \ together with the groups homomorphic to $\mathcal{I}\mathcal{H}a(
n) $ that are given in Appendix A of \cite{Low12}.\ \ 

The projective representations of $\mathcal{I}\mathcal{H}a( n) $
have been studied in \cite{Low12}.\ \ However the full projective
representations of the group $\mathcal{I}\mathcal{H}\mathcal{A}a(
n) $ has not yet been studied.\ \ It can be seen from the Lie algebra
commutations relations in Appendix B that the $M_{i,j}$ generators
of $\mathcal{A}( m) $ result from the contraction of $M_{a,b}$ that
physically are generators of\ \ force-power-stress and transform
as an (0,2) tensor under the Lorentz group.\ \ However, in the limit,
the $M_{i,j}$ generators couple only to spin generators $J_{i,j}$.\ \ This
may provide a definitive test of the theory outlined.\ \ This group
requires further investigation.\ \

\appendix\label{clg}\label{sss}\label{mk}\label{odc}\label{aff}\label{svn}\label{aga}\label{zdsp}\label{autgpga}\label{opera}\label{IHAa}

\end{document}